\documentclass[aps,prc,twocolumn,floatfix,nofootinbib,10pt]{revtex4-2}
\usepackage{amsmath,amssymb,graphicx,bm,hyperref}
\usepackage{bbm}
\usepackage{color}
\usepackage[normalem]{ulem}
\usepackage{graphicx}
\usepackage{epsfig}
\usepackage{epstopdf}
\usepackage{enumitem}

\DeclareMathOperator{\atan}{arctan}
\newcommand{\adag}[1]{a^\dagger_{#1}}
\newcommand{\aop}[1]{a^{\vphantom{\dagger}}_{#1}}

\newcommand{\vek}[1]{\bm{\mathrm{#1}}}

\newcommand{\down}{\mathord{\downarrow}}

\newcommand{\Fig}[1]{Fig.~\ref{#1}}

\newcommand{\HFB}{\text{HFB}}
\newcommand{\Hop}{\hat{H}}
\newcommand{\interaction}{\text{int}}

\newcommand{\kF}{k_{\text{F}}}
\newcommand{\EF}{E_{\text{F}}}
\newcommand{\TF}{T_{\text{F}}}
\newcommand{\kB}{k_\text{B}}
\newcommand{\kv}{\vek{k}}
\newcommand{\meanfield}{\text{mf}}
\newcommand{\Nop}{\hat{N}}
\newcommand{\Psidag}[1]{\Psi^\dagger_{#1}}
\newcommand{\Psiop}[1]{\Psi^{\vphantom{\dagger}}_{#1}}
\newcommand{\pv}{\vek{p}}
\newcommand{\Qv}{\vek{Q}}

\newcommand{\qv}{\vek{q}}

\newcommand{\up}{\mathord{\uparrow}}

\begin{document}

\title{Hartree shift and pairing gap in ultracold Fermi gases in the framework of low-momentum interactions}
\author{Michael Urban} \email{michael.urban@ijclab.in2p3.fr}
\affiliation{Universit\'e Paris-Saclay, CNRS-IN2P3, IJCLab, 91405
  Orsay, France} \author{S. Ramanan} \email{suna@physics.iitm.ac.in}
\affiliation{Department of Physics, Indian Institute of Technology
  Madras, Chennai - 600036, India}
\begin{abstract}
    In this paper we consider a two-component gas of fermions on the BCS side of the BCS-BEC crossover at zero temperature. 
    We use a momentum dependent interaction that reproduces the $s$-wave scattering phase shifts of a contact interaction up to a momentum cutoff that is scaled with the Fermi momentum.
    Using a diagrammatic formulation of Bogoliubov many-body perturbation theory, suitably augmented by self-consistency conditions, we obtain the Hartree shift and the pairing gap to third order.
    In the weak-coupling regime, our results are not only well-converged but also agree with the well-established Gor'kov-Melik-Barkhudarov corrections for the gap and the Galitskii result for the Hartree shift. 
    Near the unitary regime, our results for the Nambu-Gor'kov self-energy are less converged,
    but there is still reasonable agreement with experiments as well as with quantum Monte-Carlo results. Perspectives for improvements and applications of this approach to neutron matter are discussed.
\end{abstract}
\maketitle
\section{Introduction}
\label{sec:intro}

Ultracold gases of atoms provide a remarkable laboratory for studying strongly correlated systems, that allow for the experimental simulation of diverse complex systems as far ranging as condensed matter and nuclear systems at the atomic and subatomic scales to astrophysical systems. 

The focus of this work is an ultracold gas of fermions such as, e.g., a gas of $^6\text{Li}$ atoms. 
These systems are dilute enough that the interaction is completely specified by the scattering length $a$ (effective range $r_e = 0$).
However, the scattering length can be tuned via Feshbach resonances passing through the so called unitary regime, where $a \to \infty$ and the only remaining length scale is given by the inverse of the Fermi momentum, $\kF^{-1}$.
As a result, the energy per particle of the unitary gas is $E/N = \xi \EF$, where $\EF$ is the Fermi energy, and the Bertsch parameter $\xi\approx 0.37$~\cite{Ku2012,Horikoshi2017} is a universal constant.
By varying $a$ and hence the dimensionless parameter $(\kF a)^{-1}$ from negative to positive values, one can study the crossover from a BCS superfluid to a Bose-Einstein condensate (BEC) of bound dimers~\cite{Calvanese2018}.

Large $s$-wave scattering lengths are inherent to nuclear systems.
For example, the (spin-singlet) neutron-proton and neutron-neutron scattering lengths $a_0(np) \approx -23 \, \text{fm}$ and $a_0(nn) \approx -18 \, \text{fm}$ are much larger than the interaction range. 
It is in this context that Bertsch proposed the unitary Fermi gas as a model for a gas of neutrons at low density~\cite{Baker1999}. 
While the effective range in the neutron gas cannot be neglected, the simplicity of the atomic interactions provides a useful platform for testing various theoretical approaches.
In analogy with neutron matter, we will restrict ourselves to the BCS side of the crossover, i.e., $(\kF a)^{-1} < 0$.

The pairing gap and the normal self-energy correction (Hartree shift) are of fundamental importance as they describe the properties of the lowest fermionic quasiparticle excitations.
These can be experimentally studied, e.g., with radiofrequency spectroscopy \cite{Stewart2008}. 
The pairing gap is particularly sensitive to the approximation used and it is well known that the gap computed in the simplest BCS mean-field theory is far from satisfactory~\cite{Gorkov1961,Gezerlis2008,Calvanese2018,Biss2022}.
Even in the weak-coupling limit, particle-hole fluctuations reduce the gap by more than 50\%, which is the famous Gor'kov-Melik-Barkhudarov (GMB) correction \cite{Gorkov1961}. 
The generalization of this effect to the BCS-BEC crossover region was achieved only recently within a sophisticated diagrammatic framework \cite{Pisani2018}.

In a recent paper~\cite{Urban2021}, we constructed an effective low-momentum interaction for ultracold atoms. 
This means that, in contrast to the common practice in ultra-cold systems, the cutoff $\Lambda$ was kept finite (see Sec. \ref{sec:interaction} for more details).
As we showed in Ref. \cite{Ramanan2018} in the case of neutron matter, scaling the cutoff $\Lambda$ with $\kF$ allows one to include beyond mean field corrections, in particular screening, without resumming ladder diagrams in the vertices, whereas the diagrams necessary to describe these effects with the usual regularization procedure are extremely involved \cite{Pisani2018}.
In Ref.~\cite{Urban2021}, the ground state energy of the ultracold Fermi gas was computed within the Bogoliubov many body perturbation theory (BMBPT) up to third order.
However, we did not compute higher order corrections to the gap and the Hartree shift.
Therefore, in this work, we focus on the perturbative corrections to the self-energy (Hartree shift and the pairing gap) using the Nambu-Gor'kov formalism. In addition, in the present work, we also improve the perturbation expansion scheme as compared to our previous work~\cite{Urban2021}, namely, instead of adding corrections on top of the Hartree-Fock-Bogoliubov (HFB) ground state, we start from a state with a reduced gap, determined self-consistently.

Our work is organized as follows: in Sec.~\ref{sec:interaction}, we briefly recall the low-momentum effective separable interaction of Ref.~\cite{Urban2021} that is also used in this work, while Sec.~\ref{sec:formalism_SF} outlines the essentials of diagrammatic BMBPT with the Nambu-Gor'kov formalism.
The improved expansion scheme is then introduced in Sec.~\ref{sec:selfconsistent}.
In Sec.~\ref{sect:summary_outlook}, we summarize our findings and present our outlook.
Some more technical details are given in the appendix.

\section{Low-momentum interactions}
\label{sec:interaction}
For completeness, let us briefly summarize the set-up of the low-momentum interaction for cold atoms in Ref.~\cite{Urban2021}.
The starting point is a separable interaction in the $s$-wave, 
\begin{equation}
  V(q,q') = g F(q) F(q')\,,
  \label{Vseparable}
\end{equation}
where $g$ is the coupling constant, $q$ the relative momentum, and $F(q)$ the form factor.
In the field of cold atoms, such an interaction has been formally employed to explain the regularization of the contact interaction \cite{Pieri2000}.
Taking a sharp cutoff $\Lambda$ in momentum space, i.e., a form factor $F(q) = \theta(\Lambda-q)$, one can show that $g$ is related to the scattering length $a$ and the cutoff $\Lambda$ by (we set $\hbar=1$ throughout this article)
\begin{equation}
    \frac{1}{g} = \frac{m}{4\pi a}-\frac{m\Lambda}{2\pi^2}.
    \label{eq:RG}
\end{equation}
Using this relation to express $g$ in terms of $a$, one can finally take the limit $\Lambda\to\infty$ and obtains in this way, e.g., the regularized gap equation \cite{SadeMelo1993}.
An obvious reason why one wants to take this limit is that the form factor $F(q)=\theta(\Lambda-q)$ leads to an effective range $r_e=4/(\pi\Lambda)$, whereas the atom-atom interaction has practically zero range.

However, keeping the momentum cutoff $\Lambda$ finite can also have some advantages, as it can make perturbation theory better convergent \cite{Bogner2005}.
Also, one would expect that it should be possible to describe a system at low density and temperature in terms of only low-momentum degrees of freedom (see also \cite{Schwenk2003,Taillat2025} for an even more drastic truncation, limiting the active space to a small momentum shell around $\kF$).
In practice, we will always scale the cutoff $\Lambda$ with the Fermi momentum $\kF$, as in our previous works~\cite{Ramanan2018,Urban2020,Urban2021,Palaniappan2023,Palaniappan2025}.
In this way, we recover automatically the lowest-order in $\kF a$ of the Hartree shift in the dilute limit, since for $\Lambda\to 0$, the coupling constant $g$ approaches the value $4\pi a/m$, cf. Eq.~\eqref{eq:RG}.
The problem of the effective range can be avoided also in a different way than by taking the limit $\Lambda\to\infty$, namely, by choosing a more complicated momentum dependence of the form factor $F(q)$.

A contact interaction with scattering length $a$ is characterized by scattering phase shifts $\delta(q) = \atan(-qa)$.
We therefore want to construct an interaction that reproduces these phase shifts at momenta $q < \Lambda$. In practice, we write the phase shifts as
\begin{equation}
  \delta(q) = R(q/\Lambda) \atan (-qa)\,,
  \label{phaseshiftcontact}
\end{equation}
where
$R(q/\Lambda)$ is a smooth regulator function, such as, e.g., $e^{-(q/\Lambda)^{2n}}$ (in our numerical calculations, we take $n=10$).
Using the formula by Tabakin~\cite{Tabakin1969}, the diagonal elements of the interaction and subsequently the coupling constant and form factor are determined using 
\begin{equation}
   V(q,q) = -4\pi \frac{\sin\delta(q)}{mq} \exp\left(\frac{2}{\pi}\,
   \mathcal{P}\!\!\int_0^\infty\!\!\! dq'\, \frac{q'
     \delta(q')}{q^2-q^{\prime\,2}}\right)\,,
   \label{inversescattering}
\end{equation}
and
\begin{equation}
  g = V(0,0)\,,\quad F(q) = \sqrt{V(q,q)/g}\,.
  \label{gF}
\end{equation}

\section{Self-energy in BMBPT}
\label{sec:formalism_SF}
\subsection{Nambu-Gorkov formalism}
A Fermi gas with two spin states $\up$ and $\down$ and interaction only between opposite spins is described by the following second quantized Hamiltonian, 
\begin{multline}
  \Hop = \sum_{\pv\sigma} \frac{p^2}{2m} \adag{\pv\sigma}
  \aop{\pv\sigma}\\ + \sum_{\Qv \qv\qv'}
  V(\qv,\qv') \adag{\frac{\Qv}{2}+\qv\up} \adag{\frac{\Qv}{2}-\qv\down}
  \aop{\frac{\Qv}{2}-\qv'\down} \aop{\frac{\Qv}{2}+\qv'\up}\,,
  \label{hamiltonian}
\end{multline}
where $\aop{\pv\sigma}$ is the annihilation operator of a particle with momentum $\pv$ and spin projection $\sigma$ and the sum is a short-hand for integrations over the relevant momenta.
In the case of an $s$-wave interaction, the matrix element $V(\qv,\qv') = V(q,q')$ does not depend on the directions of the momenta.

The starting point of the Nambu-Gor'kov formalism is to consider the two-component field operators (Sec.7.2 of ~\cite{Schrieffer})
\begin{equation}
  \Psiop{\kv} = \begin{pmatrix}
    \Psiop{\kv 1}\\ \Psiop{\kv 2}
  \end{pmatrix}
  = \begin{pmatrix}
    \aop{\kv\up}\\ \adag{-\kv\down}
  \end{pmatrix}\,.
\end{equation}
Then, if we write $\qv = \tfrac{\kv+\pv'}{2}$, $\qv'=\tfrac{\kv'+\pv}{2}$, and $\Qv=\kv-\pv'=\kv'-\pv$, the interaction term of the Hamiltonian Eq.~\eqref{hamiltonian} becomes
\begin{multline}
  \Hop_{\interaction}
  = -\sum_{\kv \kv' \pv \pv'} V\big(\tfrac{\kv+\pv'}{2},\tfrac{\kv'+\pv}{2}\big)\delta_{\kv+\pv,
\kv' + \pv'}\\ 
\times {:}\Psidag{\kv{1}}\Psidag{\pv{2}}\Psiop{\pv'{2}}\Psiop{\kv'{1}}{:}\,,
  \label{HintNambu}
\end{multline}  
where colons indicate normal ordering (moving $\aop{}$ operators to
the right and $\adag{}$ operators to the left).
Following~\cite{Schrieffer}, we add and subtract the following mean field terms
\begin{align}
  \Hop_{\meanfield} &= \sum_{\pv\sigma} U_{\pv} \adag{\pv\sigma}\aop{\pv\sigma}
  +\sum_{\pv} \Delta_{\pv}
    (\adag{\pv\up}\adag{-\pv\down}+\aop{-\pv\down}\aop{\pv\up})\nonumber\\
    &= \sum_{\pv} {:}\Psidag{\pv}\begin{pmatrix}U_{\pv}&\Delta_{\pv}\\ \Delta_{\pv}&-U_{\pv}\end{pmatrix}\Psiop{\pv}{:}
    \equiv \sum_{\pv}{:}\Psidag{\pv}H_{\meanfield}(\pv)\Psiop{\pv}{:}
    \label{eq:Hmeanfield}
\end{align}
from the Hamiltonian (assuming that $\Delta_{\pv}$ is real) and write
\begin{equation}
  \Hop'_0 = \Hop_0 + \Hop_{\meanfield} - \mu\Nop\,,\quad
  \Hop'_{\interaction} = \Hop_{\interaction}-\Hop_{\meanfield}\,,
\end{equation}
where $\mu$ is the chemical potential.
Assuming that $U_{\vek{p}}$ and $\Delta_{\vek{p}}$ are independent of the direction of $\vek{p}$, we will drop the vector notation for their indices.
BMBPT consists in considering $\Hop'_0$ as the ``free'' hamiltonian and $\Hop'_{\interaction}$ as the perturbation.
$\Hop'_0$ can be written as
\begin{equation}
    \Hop'_0 = \sum_{\pv} {:}\Psidag{\pv} 
      \begin{pmatrix}
        \xi_p& \Delta_p\\ \Delta_p &-\xi_p
      \end{pmatrix}
      \Psiop{\pv}{:}\equiv\sum_{\pv} {:}\Psidag{\pv}H'_0(p)
      \Psiop{\pv}{:}\,,
\end{equation}
where
\begin{equation}
    \xi_p = \frac{p^2}{2m}+U_p-\mu\,.\label{eq:HFB-xi}
\end{equation}
The eigenvalues and eigenvectors of the matrix $H'_0(p)$ are
\begin{equation}
  E_p = \sqrt{\xi_p^2+\Delta_p^2}\,,\quad
\begin{pmatrix}u_p\\v_p\end{pmatrix} 
= \frac{1}{\sqrt{2E_p}}
\begin{pmatrix}\sqrt{E_p+\xi_p}\\ \sqrt{E_p-\xi_p}\end{pmatrix}.
\label{eq:HFB-Euv}
\end{equation}
Hence, the eigenstates of $H'_0$ are quasiparticles with energy $E_p$.

The mean-field Green's function corresponding to $\Hop'_0$ is a $2\times 2$ matrix that is given by
  \begin{gather}
    G^{(0)}(k,\omega) = \begin{pmatrix}
      G^{(0)}_{11}(k,\omega)& G^{(0)}_{12}(k,\omega)\\
      G^{(0)}_{21}(k,\omega)& G^{(0)}_{22}(k,\omega)
  \end{pmatrix},\\
    G^{(0)}_{\alpha\beta}(k,\omega) =
    \frac{A_{\alpha\beta}({k})}{\omega-E_{k}+i\eta}
    +\frac{B_{\alpha\beta}(k)}{\omega+E_{k}-i\eta}\,,\label{GwithAB}\\
    A(k) = \begin{pmatrix}
      u_k^2& u_k v_k\\u_k v_k& v_k^2
    \end{pmatrix},\quad 
    B(k) = \begin{pmatrix}
      v_k^2& -u_k v_k\\-u_k v_k& u_k^2
    \end{pmatrix},\label{matrixAB}
  \end{gather}
where the limit $\eta\to 0^+$ is understood. 

\begin{figure}
  \includegraphics[scale=0.94]{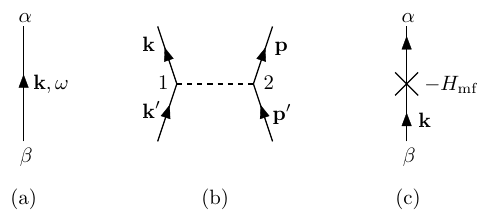}
  \caption{\label{fig:feynmanrules} Elements for Feynman diagrams:
  (a) $G^{(0)}_{\alpha\beta}(k,\omega)$, 
  (b) $-V(\tfrac{\kv+\pv'}{2},\tfrac{\kv'+\pv}{2})$,
  (c) $-H_{\meanfield,\alpha\beta}(k)$.}
\end{figure}
We then obtain Feynman rules (in momentum space) for the computation of the (interacting) Green's function $G$.
The basic elements are shown in Fig.~\ref{fig:feynmanrules}, more details are given in Appendix \ref{app:feynmanrules}.

Notice that, since we are using matrix propagators, a line can represent normal and anomalous Green's functions, depending on the Nambu indices.
For lines with Nambu index 2, one has to remember that particles and holes are interchanged.
This explains also the momentum arguments in the interaction vertices that look surprising at first glance but which can be understood if one remembers that the outgoing and incoming lines with Nambu index $2$ and momenta $\pv$ and $\pv'$ describe in reality incoming and outgoing particles with momenta $-\pv'$ and $-\pv$, respectively.

\begin{figure}
  \includegraphics[scale=0.94]{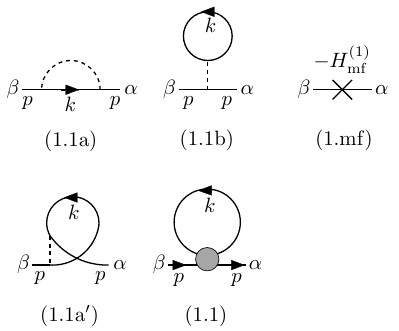}
  \caption{\label{fig:sigma1} (1.1a), (1.1b), and (1.mf) First-order self-energy diagrams.
  Diagram (1.1a) has been drawn differently in (1.1a$'$) to show that (1.1a) and (1.1b) can be compactly represented by one generic diagram denoted (1.1).
  (For simplicity, we denote the loop momentum as $k$ instead of $\vek{k},\omega$ in the figures.)}
\end{figure}
\subsection{Self-energy at first order and HFB}

At first order, there are three self-energy diagrams shown in the upper row of \Fig{fig:sigma1}.
To illustrate the Feynman rules, let us compute step by step the diagrams (1.1a) and (1.1b).
Since only two particles with opposite spins can interact, it is evident that diagram (1.1a) contributes only to the non-diagonal (in Nambu indices) matrix elements of the self-energy, while diagram (1.1b) contributes to the diagonal ones.
When we consider $\Sigma_{12}$, the right vertex of diagram (1.1a) has Nambu index 1 and the left vertex has Nambu index 2.
Let us denote the momentum of the external lines $\pv$ (notice that at first order, the self-energy does not depend on the energy) and the energy and momentum of the intermediate propagator $\omega$ and $\kv$ (for brevity simply labeled by $k$ in the figure). 
Then, the potential corresponding to the interaction line is $V(\pv,\kv)$. 
Therefore, the expression corresponding to diagram (1.1a) reads (we drop the superscript $(0)$ in the mean-field propagator $G^{(0)}$ to simplify the notation)
\begin{align}
  \Sigma^{\text{(1.1a)}}_{12}(p) &= 
  -i \int\!\frac{d^3k\,{d\omega}}{(2\pi)^4} V(\pv,\kv)
  G_{12}(k{,\omega})\nonumber\\
  &=-\int\!\frac{d^3k}{(2\pi)^3} V(\pv,\kv) u_k v_k \nonumber\\
  &{= -g F(p)\int\! \frac{dk}{2\pi^2} k^2 F(k) u_k v_k}\,.
\end{align}
Repeating the same steps for $\Sigma^{\text{(1.1a)}}_{21}$, one finds the same result.

Diagram (1.1b) contributes to $\Sigma_{11}$ and $\Sigma_{22}$.
Let us consider $\Sigma_{11}$. In this case, the propagator in the closed loop (which gives rise to a minus sign) must have Nambu indices 22 and the interaction contributes a factor of $V(\frac{\pv+\kv}{2},\frac{\kv+\pv}{2})$.
Putting everything together, we find
\begin{align}
  \Sigma^{\text{(1.1b)}}_{11}(p) &=
  i\int\!\frac{d^3k\,d\omega}{(2\pi)^4}
  V(\tfrac{\pv+\kv}{2},\tfrac{\pv+\kv}{2})G_{22}(k,\omega) e^{-i\eta \omega}\nonumber\\
  &=\int\!\frac{d^3k}{(2\pi)^3} V(\tfrac{\pv+\kv}{2},\tfrac{\pv+\kv}{2}) v^2_k \nonumber\\
  &= \int\! \frac{dk}{2\pi^2}\,k^2 \bar{V}(p,k) v_k^2\,,
\end{align}
where $\bar{V}(k,p)$ in the last line denotes the average of the matrix element in the second line over the angle between $\pv$ and $\kv$.
Similarly, one finds that $\Sigma^{\text{(1.1b)}}_{22}(k) = -\Sigma^{\text{(1.1b)}}_{11}(k)$.

The usual HFB approach is to determine $U_p$ and $\Delta_p$ such that $\Sigma^{\text{(1.1a)}}$ and $\Sigma^{\text{(1.1b)}}$ are exactly cancelled by the self-energy contributions generated by the term $-\Hop_{\meanfield}$ in the perturbation hamiltonian $\Hop'_{\interaction}$, represented by the diagram (1.mf) \cite{Schrieffer}.
If we denote the first-order self-energy without the $-H_{\meanfield}$ contribution by $\Sigma^{(1)}$ (i.e., $\Sigma^{(1)} = \Sigma^{(\text{1.1a})}+\Sigma^{(\text{1.1b})}$) and the total first-order self-energy including the $-H_{\meanfield}$ term by $\Sigma'^{(1)}$, this can be written as
\begin{equation}
  \Sigma'^{(1)}(p)=\Sigma^{(1)}(p) - H^{(1)}_{\meanfield}(p)
   = 0\,.\label{eq:HFB-Hmf}
\end{equation}
Hence, the matrix elements of $H_{\meanfield}$ must be
\begin{equation}
  U_p^{(1)} = \Sigma^{(\text{1.1b})}_{11}(p)\,,\quad
  \Delta_p^{(1)} = \Sigma^{(\text{1.1a})}_{12}(p)\,.
  \label{eq:HFB-UDelta}
\end{equation}
Here, we have denoted $H_{\meanfield}$, $U_p$, and $\Delta_p$ as $H^{(1)}_{\meanfield}$, $U^{(1)}_p$, and $\Delta^{(1)}_p$ for reasons that will be explained in Sec.~\ref{sec:selfconsistent}.

\begin{figure}[b!]
  \vspace*{3mm}
  \includegraphics[scale=0.94]{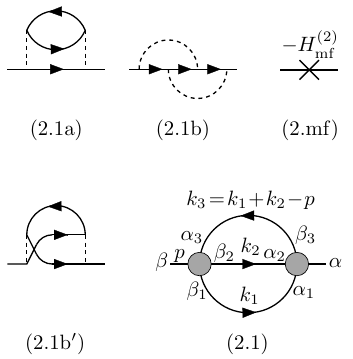}
  \caption{\label{fig:sigma2} (2.1a), (2.1b), and (2.mf) Diagrams for the second-order self-energy $\Sigma^{(2)}$. 
  (2.1b$'$) Alternative way of drawing diagram (2.1b). 
  (2.1) Generic diagram combining (2.1a) and (2.1b), showing the labeling used in the calculation.}
\end{figure}
\begin{figure}[b!]
  \vspace*{3mm}
  \includegraphics[scale=0.94]{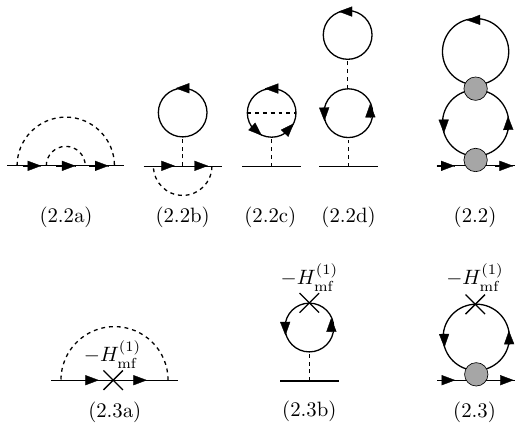}
  \caption{\label{fig:sigma2-cancel} Second-order self-energy diagrams that cancel as a consequence of Eq.~\eqref{eq:HFB-Hmf}.}
\end{figure}
\subsection{Self-energy at second order}\label{sec:Sigma2}
The second-order diagrams that contribute to $\Sigma$ are the diagrams (2.1a) and (2.1b) shown in \Fig{fig:sigma2}.
The diagram (2.1b) can also be drawn differently as shown in (2.1b$'$).
Drawn in this way, it is easy to see that both diagrams have the same general form depicted in diagram (2.1).
The diagram denoted (2.mf) does not exist if one builds the perturbation theory expansion on top of the HFB ground state, but it will appear once we include higher-order corrections more self-consistently as we will discuss in Sec.~\ref{sec:selfconsistent}.
Furthermore, there are other second-order diagrams (2.2a)-(2.2d) and (2.3a)-(2.3b) shown in \Fig{fig:sigma2-cancel}, which again can be combined into more generic diagrams denoted (2.2) and (2.3).
But these diagrams contain first-order self-energy insertions and hence their contribution is zero due to Eq.~\eqref{eq:HFB-Hmf}.

Let us turn to the evaluation of the two diagrams of the class (2.1). Denoting the external momentum and energy by $p$ and $\omega$, respectively, both diagrams contain an integral of the form
\begin{widetext}
\begin{align}
  I^{(2)}_{\alpha_1\beta_1\alpha_2\beta_2\alpha_3\beta_3}
    (\omega,k_1,k_2,k_3)
   &= (-i)^2 \int\frac{d\omega_1}{2\pi}\int\frac{d\omega_2}{2\pi}
  G_{\alpha_1\beta_1}(k_1,\omega_1)G_{\alpha_2\beta_2}(k_2{,\omega_2})
  G_{\alpha_3\beta_3}(k_3,{\omega_1}+{\omega_2}-{\omega})
  \nonumber\\
  &= -\frac{A_{\alpha_1\beta_1}(k_1)A_{\alpha_2\beta_2}(k_2)
    B_{\alpha_3\beta_3}(k_3)}{{\omega}-E_{k_1k_2k_3}+i\eta}
  - \frac{B_{\alpha_1\beta_1}(k_1)B_{\alpha_2\beta_2}(k_2)
    A_{\alpha_3\beta_3}(k_3)}{{\omega}+E_{k_1k_2k_3}-i\eta}\,,
  \label{integral2}
\end{align}
with $E_{k_1k_2k_3} = E_{k_1}+E_{k_2}+E_{k_3}$. In each of the diagrams (2.1a) and (2.1b), all internal Nambu indices $\alpha_1\dots\beta_3$ are completely determined by the external indices $\alpha$ and $\beta$.
Using the notation $\bar{\alpha} = 3-\alpha$ (i.e., $\bar{1}=2$ and $\bar{2}=1$) and $\kv_3 = \kv_1+\kv_2-\pv$, we obtain for diagram (2.1a)
\begin{equation}
  \Sigma^{\text{(2.1a)}}_{\alpha\beta}(p,\omega) = -\int \frac{d^3k_1}{(2\pi)^3}
  \int\frac{d^3k_2}{(2\pi)^3}
  \left[V(\tfrac{\kv_1+\kv_3}{2},\tfrac{\kv_2+\pv}{2})\right]^2
  I^{(2)}_{\alpha\beta\bar{\alpha}\bar{\beta}\bar{\beta}\bar{\alpha}}
    (\omega,k_1,k_2,k_3)\,.
  \label{Sigma2ageneral}
\end{equation}
The negative sign comes from the closed Fermion loop. In diagram (2.1b), there is no closed Fermion loop and we obtain
\begin{equation}
  \Sigma^{\text{(2.1}b)}_{\alpha\beta}(p,\omega) = \int \frac{d^3k_1}{(2\pi)^3}
  \int\frac{d^3k_2}{(2\pi)^3}
   V(\tfrac{\kv_1+\kv_3}{2},\tfrac{\kv_2+\pv}{2})V(\tfrac{\kv_2+\kv_3}{2},\tfrac{\kv_1+\pv}{2})
   I^{(2)}_{\alpha\bar{\beta}\bar{\alpha}\beta\bar{\beta}\bar{\alpha}}
  (\omega,k_1,k_2,k_3)\,.
  \label{Sigma2bgeneral}      
\end{equation}
With Eq.~\eqref{matrixAB} and \eqref{integral2}, it is now straight-forward to get the explicit expressions for $\Sigma_{11}$ and $\Sigma_{12}$
\begin{align}
    \Sigma^{\text{(2.1a)}}_{11}(p{,\omega}) &= \int \frac{d^3k_1}{(2\pi)^3}
    \int\frac{d^3k_2}{(2\pi)^3}
    \left[V(\tfrac{\kv_1+\kv_3}{2},\tfrac{\kv_2+\pv}{2})\right]^2
      \Bigg(\frac{u_{k_1}^2v_{k_2}^2u_{k_3}^2}{\omega-E_{k_1k_2k_3}+i\eta}
      +\frac{v_{k_1}^2u_{k_2}^2v_{k_3}^2}{\omega+E_{k_1k_2k_3}-i\eta}\Bigg)\,,\\
    \Sigma^{\text{(2.1b)}}_{11}(p,\omega) &= \int \frac{d^3k_1}{(2\pi)^3}
    \int\frac{d^3k_2}{(2\pi)^3}
     V(\tfrac{\kv_1+\kv_3}{2},\tfrac{\kv_2+\pv}{2})
     V(\tfrac{\kv_2+\kv_3}{2},\tfrac{\kv_1+\pv}{2})
     \Bigg(-\frac{u_{k_1}v_{k_1}u_{k_2}v_{k_2}u_{k_3}^2}
          {{\omega}-E_{k_1k_2k_3}+i\eta}
          -\frac{u_{k_1}v_{k_1}u_{k_2}v_{k_2}v_{k_3}^2}
          {\omega+E_{k_1k_2k_3}-i\eta}\Bigg)\,,\\
    \Sigma^{\text{(2.1a)}}_{12}(p,\omega) &= \int \frac{d^3k_1}{(2\pi)^3}
    \int\frac{d^3k_2}{(2\pi)^3}
    \left[V(\tfrac{\kv_1+\kv_3}{2},\tfrac{\kv_2+\pv}{2})\right]^2
    \Bigg(-\frac{u_{k_1}v_{k_1}u_{k_2}v_{k_2}u_{k_3}v_{k_3}}
         {{\omega}-E_{k_1k_2k_3}+i\eta}
         +\frac{u_{k_1}v_{k_1}u_{k_2}v_{k_2}u_{k_3}v_{k_3}}
         {{\omega}+E_{k_1k_2k_3}-i\eta}\Bigg)\,,\label{Sigma2a12}\\
    \Sigma^{\text{(2.1b)}}_{12}(p,\omega) &= \int \frac{d^3k_1}{(2\pi)^3}
    \int\frac{d^3k_2}{(2\pi)^3}
     V(\tfrac{\kv_1+\kv_3}{2},\tfrac{\kv_2+\pv}{2})
     V(\tfrac{\kv_2+\kv_3}{2},\tfrac{\kv_1+\pv}{2})
     \Bigg(\frac{u_{k_1}^2 v_{k_2}^2u_{k_3}v_{k_3}}
         {{\omega}-E_{k_1k_2k_3}+i\eta}
         -\frac{v_{k_1}^2u_{k_2}^2u_{k_3}v_{k_3}}
         {{\omega}+E_{k_1k_2k_3}-i\eta}\Bigg)\,.\label{Sigma2b12}
  \end{align}
\end{widetext}
We will evaluate these integrals with Monte-Carlo integration with importance sampling. As in the computation of the ground state energy \cite{Urban2021}, one prefers integration variables with weight factors $uv$ or $v^2$ over those with $u^2$. 
But it is not a problem to switch the integration variables: any two of the three momenta $\kv_1, \kv_2, \kv_3$ can be used as integration variables if the third momentum is computed according to $\kv_1+\kv_2-\kv_3 = \pv$.

We have seen that at first order, there are simple relations between $\Sigma_{11}$ and $\Sigma_{22}$ and between $\Sigma_{12}$ and $\Sigma_{21}$. 
Actually, such relations exist at all orders as shown in Appendix \ref{app:symmetry-properties}, namely:
\begin{align}
    \Sigma_{22}(p,\omega) &= -\Sigma_{11}(p,-\omega)\,,\label{eq:selfenergy-symmetry1a}\\
    \Sigma_{21}(p,\omega) &= \Sigma_{12}(p,-\omega)\,.\label{eq:selfenergy-symmetry1b}\\
  \Sigma_{12}(p,\omega) &= \Sigma_{21}(p,\omega)\,.\label{eq:selfenergy-symmetry2}
\end{align}
As a consequence, it is sufficient to compute only $\Sigma_{11}$ and $\Sigma_{12}$. 
Furthermore, combining Eq.~\eqref{eq:selfenergy-symmetry1b} with Eq.~\eqref{eq:selfenergy-symmetry2} one finds that $\Sigma_{12}$ is an even function of energy. Similar symmetry relations were derived in \cite{Soma2011} for the more general case of non-uniform systems.

\begin{figure}
  \includegraphics[width=\columnwidth]{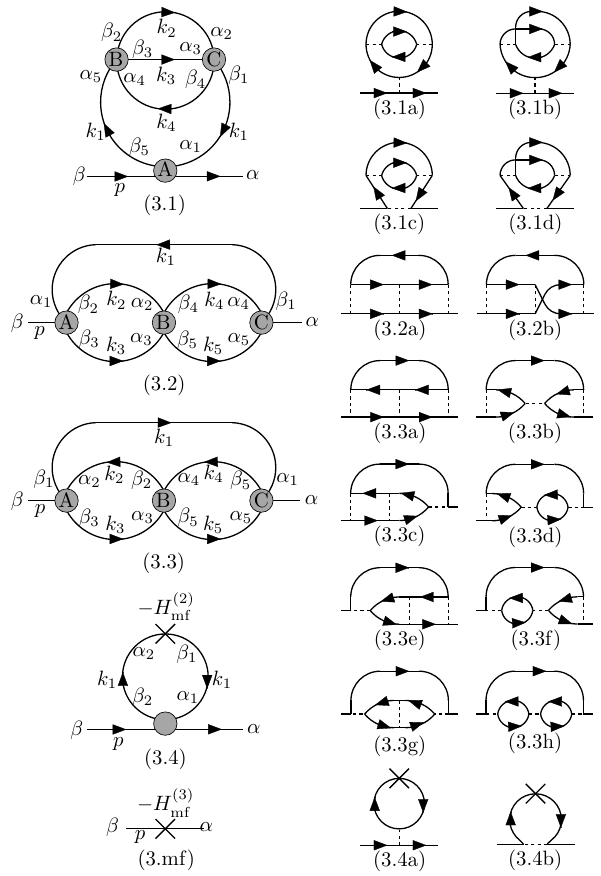}
  \caption{\label{fig:sigma3} Left column, (3.1)-(3.3): general form of the three classes of diagrams contributing to the third-order self-energy
    $\Sigma^{(3)}$ in perturbation theory on top of the HFB ground state.
    (3.4)-(3.mf): additional diagrams appearing in the more self-consistent scheme.
    Right column (3.1a)-(3.4b): corresponding individual diagrams.}
\end{figure}
\subsection{Self-energy at third order}
The third-order diagrams that contribute to $\Sigma$ are shown in
\Fig{fig:sigma3}.
If we perform the perturbation calculation on top of the HFB ground state, there are in total 14 diagrams (3.1a)-(3.3h), shown in the right column, which can be classified according to the energy integrations into three different groups (3.1)-(3.3), shown in the left column.
We do not draw diagrams with first-order self-energy insertions since they vanish because of Eq.~\eqref{eq:HFB-Hmf}, analogous to the diagrams (2.2) and (2.3) in Fig.~\ref{fig:sigma2-cancel}, as discussed in Sec.~\ref{sec:Sigma2}. The details of the calculation of the diagrams (3.1)-(3.3) are given in Appendix~\ref{app:3rdorder}.
The other diagrams of type (3.4) and (3.mf) shown in Fig.~\ref{fig:sigma3} will be discussed in Sec.~\ref{sec:selfconsistent}. 

\begin{figure*}
    \begin{center}
        \includegraphics{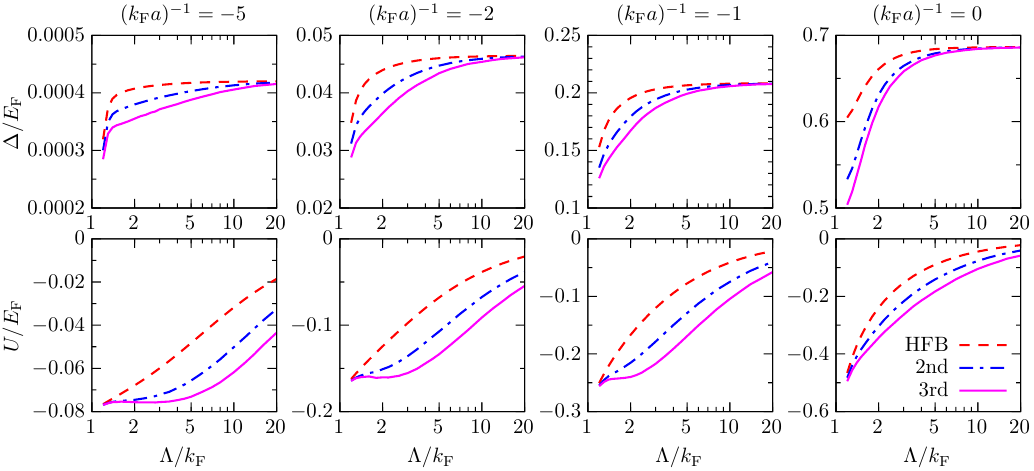}
    \end{center}
    \caption{Cutoff dependence of the gap (upper row) and mean field shift (lower row) computed within HFB+perturbation theory up to 3rd order for coupling strengths $(\kF a)^{-1}= -5,-2,-1,0$ (from left to right).
    \label{fig:cutoff-BMBPT}}
\end{figure*}
\subsection{Preliminary results}
\label{sec:perturbative-results}
In this section, we discuss some results for the normal and anomalous self-energy within diagrammatic perturbation theory, shown in \Fig{fig:cutoff-BMBPT}.
The upper row shows the pairing gap $\Delta = \Sigma_{12}$, in units of $\EF = \kF^2/(2m)$ and evaluated at $k = \kF$, $\omega = E_{\kF} (\approx \Delta_{\HFB})$.
The lower row shows the Hartree shift $U = \Sigma_{11}$, again in units of $\EF$ but evaluated at $k = \kF$, $\omega = 0$.
(For a discussion about this choice, cf. next paragraph.) It must be emphasized that beyond HFB, $U$ is strictly speaking no longer a ``Hartree shift,'' but this terminology has also been used in the previous literature \cite{Perali2002,Schirotzek2008}. 
As mentioned in Secs.~\ref{sec:intro} and \ref{sec:interaction}, we scale the cutoff $\Lambda$ of our low-momentum interaction with the Fermi momentum $\kF = (3\pi^2 n)^{1/3}$.
Results are plotted as functions of the ratio $\Lambda/\kF$ for four different interaction strengths $1/(\kF a) = -5$, $-2$, $-1$ and $0$ (unitary Fermi gas).

Before discussing the figure, let us briefly comment on the different choices of $\omega$ 
for $\Delta$ and $U$. To define the gap at $k=\kF$, $\omega=E_{\kF}$ seems reasonable as this
corresponds to the minimum of the quasiparticle energy at the HFB level. But in the case of 
$\Sigma_{11}$, remember that it is not an even function of $\omega$ (in contrast to
$\Sigma_{12}$), and $\omega=0$ is the best compromise between the two quasiparticle energies
$\pm E_{\kF}$. Furthermore, this definition of $U$ is very similar to the one of $\Sigma_0$ in 
Refs.~\cite{Pieri2004,Pisani2018}.

As one can see in Fig.~\ref{fig:cutoff-BMBPT}, $\Delta$ approaches a constant, namely its BCS
value, in the limit $\Lambda\to \infty$, while the Hartree shift disappears in this limit because the 
matrix elements of the interaction $V$ go to zero as $1/\Lambda$. For small values of $\Lambda/\kF$, 
the HFB results (red dashes) show a strong cutoff dependence and one would hope that, as it was the 
case for the energy \cite{Urban2021}, this cutoff dependence is reduced when one includes 
perturbative corrections.

For the Hartree shift $U$, this is indeed the case, at least in the weakly interacting regime 
$1/(\kF a) = -5$: when including second-order (blue dash-dot lines) and third-order (purple 
solid lines) corrections, a plateau builds up in the cutoff range 
$1.5\lesssim \Lambda/\kF \lesssim 4$. Also, we see that for $\Lambda/\kF \lesssim 2$, the 
third-order correction is much smaller than the second-order one, which shows the convergence
of the perturbative expansion. The reason why nevertheless the cutoff dependence becomes stronger again for $\Lambda/\kF\lesssim 1.5$ may be related to the neglect of induced three- and higher-body interactions (cf. discussion in Sec.~\ref{sect:summary_outlook}). However increasing the interaction to $1/(\kF a) = -2$ and 
$ -1$, the cutoff range of the plateau shrinks, and in the unitary limit, there is no 
plateau at all.

Unlike the Hartree shift $U$, the pairing gap $\Delta$ does not show any convergence even in 
the weakly coupled regime. We see that the third-order correction is as large as the 
second-order one for all cutoffs, and there is no plateau. In fact, what is expected
in the weakly coupled regime is a suppression of the gap by a factor of 
$\approx 0.45$ \cite{Gorkov1961}. This result can be obtained in a perturbative calculation, 
but for the effective in-medium interaction and not directly for the self-energy. Then,
the screened interaction is used in the highly non-linear gap equation, and by solving
the latter, the corrections to the interaction are taken into account non-perturbatively.
To achieve the same effect in the framework of the Nambu-Gorkov self-energy, an extension
of the perturbative scheme is required, as we will explain in the next section.

\section{A more self consistent scheme}
\label{sec:selfconsistent}
\subsection{Formalism}
\label{sec:selfconsistent-formalism}
Following \cite{Schrieffer}, one would compute the mean field propagators with the HFB solution, i.e., $H_{\meanfield} = H^{(1)}_{\meanfield}$ as given in Eqs. \eqref{eq:HFB-Hmf}-\eqref{eq:HFB-UDelta}.
Then, $-H_{\meanfield}$ cancels the first-order self-energy $\Sigma^{(1)}$ and diagrams containing first-order self-energy insertions or $-H_{\meanfield}$ can be discarded.
This is equivalent to building BMBPT on top of the HFB ground state, as it was done in the computation of the ground-state energy in Ref. \cite{Urban2021}.

However, since we already know that the gap will be considerably reduced compared to the HFB one due to medium-polarization effects \cite{Gorkov1961}, it may be better to use from the beginning $H_{\meanfield}$ and corresponding propagators $G$ that have the reduced gap rather than the HFB one.
This means that we have to achieve some approximate self-consistency between $U$ and $\Delta$ in $H_{\meanfield}$ and the higher-order self-energy.
A scheme using completely self-consistent Green's functions, also known as $\Phi$-derivable or Luttinger-Ward scheme, was implemented in Ref. \cite{Haussmann2007}, but this goes beyond the scope of the present work. 
Notice that also in Ref.~\cite{Pisani2018}, self-consistency for the gap was imposed.

To simplify the discussion, we will again denote the self-energy without the $-H_{\meanfield}$ contribution by $\Sigma$ and the total self-energy including the $-H_{\meanfield}$ contribution by $\Sigma'$ (i.e., $\Sigma'= \Sigma-H_{\meanfield}$). Then, the aim is to find an $H_{\meanfield}$ such that $\Sigma'\approx 0$, which is equivalent to $H_{\meanfield}\approx \Sigma$.
Obviously, except at first order, $H_{\meanfield}$ cannot cancel $\Sigma$ exactly, because $H_{\meanfield}(k)$ depends only on the momentum, while the self-energy $\Sigma(k,\omega)$ depends also on the energy.

When working with mean field Green's functions based on $H_{\meanfield}$, the best one could do would be to set $H_{\meanfield}(k)$ equal to $\Sigma(k,\omega)$ at some energy, e.g., at $\omega = E_k$ or $\omega=0$. But even this would be numerically demanding because it would require to compute $\Sigma(k, E_k)$ or $\Sigma(k,0)$ self-consistently for all $k$. 
So, we will impose only that $\Sigma'$ vanishes at one chosen momentum, e.g., at the Fermi momentum.

More precisely, we set
\begin{equation}
  U_{\kF} = \Sigma_{11}(\kF,0)\,,\quad
  \Delta_{\kF} = \Sigma_{12}(\kF, E_{\kF})\,.
  \label{eq:selfconsistent}
\end{equation}
Similarly to the discussion in Sec.~\ref{sec:perturbative-results}, the reason why we choose different energies as renormalization points for $U$ and $\Delta$ is that the mean-field propagator for $k=\kF$ has two poles, namely at $\omega=\pm E_{\kF}$, but $\Sigma_{11}$ is not an even function of $\omega$, whereas $\Sigma_{12}$ is [cf. Eqs. \eqref{eq:selfenergy-symmetry1a}-\eqref{eq:selfenergy-symmetry2}]. 
In practice, however, we checked that the results remain practically unchanged if we compute also $\Sigma_{12}$ at $\omega=0$ instead of $\omega=E_{\kF}$.

Now, let us write the perturbation as
\begin{equation}
    \Hop'_{\interaction} = \lambda \Hop_{\interaction} - \lambda \Hop^{(1)}_{\meanfield} - 
     \lambda^2 \Hop^{(2)}_{\meanfield} - 
      \lambda^3 \Hop^{(3)}_{\meanfield} - \cdots ,
\end{equation}
where $\lambda$ is a formal expansion parameter which will be set to $\lambda=1$ in the end. 
Let $\Sigma'^{(n)}$ be the self-energy contribution proportional to $\lambda^n$.
At every order $n$, there is now a contribution $-H^{(n)}_{\meanfield}$ represented by the diagram denoted ($n$.mf), while the remaining diagrams give $\Sigma^{(n)}$.
Writing Eq. \eqref{eq:selfconsistent} order by order in $\lambda$, we get
\begin{equation}
  U^{(n)}_{\kF} = \Sigma^{(n)}_{11}(\kF,0)\,,\quad
  \Delta^{(n)}_{\kF} = \Sigma^{(n)}_{12}(\kF, E_{\kF})\,,
  \label{eq:selfconsistentorder}
\end{equation}
where $U^{(n)}$ and $\Delta^{(n)}$ are the matrix elements of $H^{(n)}_{\meanfield}$. 
Since, except for $n=1$, $\Sigma'^{(n)}(k,\omega$) does not vanish at all values of $k$ and $\omega$, diagrams containing self-energy insertions of order $n\ge 2$ can still give a finite contribution.
For example, at third order, the diagram (3.4) with an insertion of $-H_{\meanfield}^{(2)}$ cancels the diagram (3.1) only approximately, unlike diagrams (2.2) and (2.3), which cancel exactly.

Equations \eqref{eq:selfconsistentorder} do not say anything about the momentum dependence of $U^{(n)}_k$ and $\Delta^{(n)}_k$.
For $n=1$, we keep it as given by Eq. \eqref{eq:HFB-UDelta}.
But for $n\ge 2$, we have to make a choice.
Namely, we choose
\begin{equation}
    U^{(n)}_k = U^{(n)}_{\kF}\,,\quad \Delta^{(n)}_k = \Delta^{(n)}_{\kF} \frac{F(k)}{F(\kF)}\,.
\label{eq:momentumdependence}
\end{equation}
This choice requires only minor modifications of the numerical implementation compared to the perturbative scheme based on HFB, because it amounts to simply replacing in Eq. \eqref{eq:HFB-xi} the chemical potential by an effective one,
\begin{equation}
    \xi_k = \frac{k^2}{2m}+U^{(1)}_k - \mu^*\,,\quad
    \mu^* = \mu - \sum_{n\ge 2} U^{(n)}_{\kF}\,,
\end{equation}
while the momentum dependence of $\Delta_k$ determined by the form factor $F(k)$ of the separable potential has the same shape as in HFB. In spite of this simple momentum dependence, the cancellation between the diagrams (3.1) and (3.4), mentioned above, works in practice quite well. From now on, we will simply write $U$, $U^{(n)}$, $\Delta$, and $\Delta^{(n)}$ as short-hand notations for $U_{\kF}$, $U^{(n)}_{\kF}$, $\Delta_{\kF}$, and $\Delta^{(n)}_{\kF}$, respectively.

The counterterm contributions to the third-order self-energy shown in diagram (3.4) can be easily obtained from
\begin{equation}
  \Sigma^{(3.4)}(p) = 
  \frac{\partial\Sigma^{(1)}(p)}{\partial \mu^*} U^{(2)}
  -\frac{\partial\Sigma^{(1)}(p)}{\partial\Delta} \Delta^{(2)}\,.
\end{equation}
The derivatives can be computed explicitly, with the results:
\begin{align}
\Sigma^{(3.4)}_{11}(p) =\,
    &\int\! \frac{dk}{2\pi^2}\, k^2 \bar{V}(p,k)\frac{\Delta_k^2}{2E_k^3}\, U^{(2)}\nonumber\\
    &-\int\! \frac{dk}{2\pi^2}\, k^2 \bar{V}(p,k)\frac{F(k)\xi_k\Delta_k}{2E_k^3}\,
      \frac{\Delta^{(2)}}{F(\kF)}\,,\\
  \Sigma^{\text{(3.4)}}_{12}(p) =\,
    &-g F(p) \int\! \frac{dk}{2\pi^2}\, k^2 \frac{F(k)\xi_k\Delta_k}{2E_k^3}\, U^{(2)}\nonumber\\
    &+g F(p) \int\! \frac{dk}{2\pi^2}\, k^2 \frac{[F(k)\xi_k]^2}{2E_k^3}\, 
      \frac{\Delta^{(2)}}{F(\kF)}\,.
\end{align}

Furthermore, in our numerical implementation, we make the assumption that the density is unaffected by the momentum and energy dependence of the second- and third-order self-energy contributions, thereby greatly simplifying the calculation, as only the gap equation has to be solved self-consistently. This is a point that should be improved in future work.

At third order, the algorithm can be summarized as follows: 
\begin{enumerate}
    \item We need a function that computes for given $\kF$ and $\Delta$:
    \begin{enumerate}
        \item the effective chemical potential $\mu^*$ and the first-order mean field $U_k^{(1)}$ such that
        \begin{equation}
            \frac{\kF^3}{3\pi^2} = \int\! \frac{d^3k}{(2\pi)^3}\, v_k^2\,.
        \end{equation}
        where the $u$ and $v$ factors are computed with the given $\Delta$ and not with $\Delta^{(1)}$.
        \item the first-order gap $\Delta^{(1)}$.
        \item the second-order mean field $U^{(2)}$ and gap $\Delta^{(2)}$.
        \item the third-order gap $\Delta^{(3)}$ [diagrams (3.1)-(3.4), using in $\Sigma_{12}^{(3.4)}$ the values of $U^{(2)}$ and $\Delta^{(2)}$ computed in step (c)].
    \end{enumerate}
    \item Find the self-consistent value of $\Delta$ by solving the non-linear equation
      \begin{equation}
        \Delta = \Delta^{(1)}+\Delta^{(2)}+\Delta^{(3)}\,.
        \label{eq:selfconsistent-Delta}
      \end{equation}
    \item Compute $U^{(3)}$ [diagrams (3.1)-(3.4)] to obtain the mean field shift
    \begin{equation}
        U = U^{(1)}+U^{(2)}+U^{(3)}\,.
        \label{eq:selfconsistent-U}
    \end{equation}
\end{enumerate}
\begin{figure}
    \begin{center}
        \includegraphics{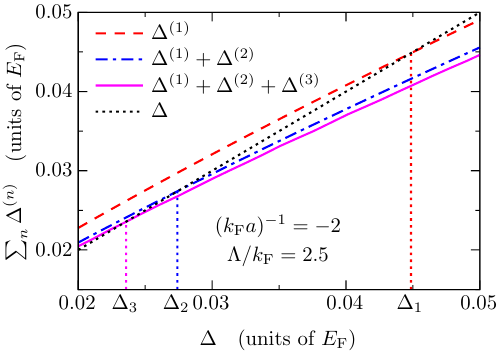}
    \end{center}
    \caption{Graphical solution of Eq.~\eqref{eq:selfconsistent-Delta} by plotting the rhs, truncated at different orders, and the lhs ($=\Delta$, black dotted line) as functions of $\Delta$. The solutions at the different orders (indicated by the vertical dotted lines) are where the curves intersect the black dotted line.}
    \label{fig:self-consistency}
\end{figure}
\begin{figure*}
    \begin{center}
        \includegraphics{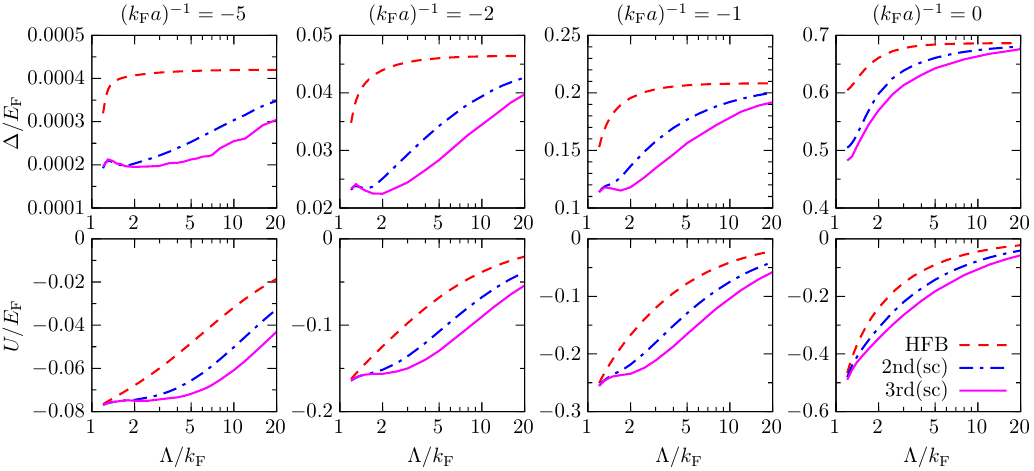}
    \end{center}
    \caption{Cutoff dependence of the gap (upper row) and mean field shift (lower row) computed within the selfconsistent scheme for coupling strengths $(\kF a)^{-1}= -5,-2,-1,0$ (from left to right).}
    \label{fig:cutoff-selfcon}
\end{figure*}
\subsection{Illustration of the self-consistent procedure}
\label{sec:illustration}
Let us illustrate the procedure defined by Eq.~\eqref{eq:selfconsistent-Delta} in a specific case. We choose $(\kF a)^{-1}=-2$, corresponding to a relatively weak coupling where one expects perturbation theory to converge.

The HFB solution corresponds to solving $\Delta=\Delta^{(1)}$. $\Delta^{(1)}$ as function of $\Delta$ is shown as the red dashed line in Fig.~\ref{fig:self-consistency}, while $\Delta$ is shown as the black dotted line. So, the HFB solution for $\Delta$, which we will denote $\Delta_1$, is the intersection of these two lines. At second order, one has to solve $\Delta=\Delta^{(1)}+\Delta^{(2)}$. $\Delta^{(1)}+\Delta^{(2)}$ is shown as the blue dash-dotted line. As one can see, although $\Delta^{(2)}$ is less than a $10\%$ correction to $\Delta^{(1)}$, the second-order solution $\Delta_2$ (intersection of the blue dash-dotted and the black dotted lines) is much smaller than the HFB solution. At third order, this effect is even more pronounced: $\Delta^{(1)}+\Delta^{(2)}+\Delta^{(3)}$ (purple solid line) is very close to $\Delta^{(1)}+\Delta^{(2)}$, but the third-order solution $\Delta_3$ (intersection of the purple solid and the black dotted lines) is still significantly smaller than $\Delta_2$, simply because the slope of the curve $\Delta^{(1)}+\Delta^{(2)}+\Delta^{(3)}$ as a function of $\Delta$ is close to one. 

In this way, it is possible to understand the strong reduction of the gap in the weak-coupling regime predicted by GMB \cite{Gorkov1961}, whereas when computing perturbative corrections with $\Delta$ fixed to its HFB value, one finds only a moderate reduction of the gap. This is why the self-consistency is so important.

\subsection{Convergence and cutoff dependence}
\label{sec:cutoff-dependence}
In the preceding example, the third-order correction was found to still be important. Therefore, it is necessary to estimate the 
contributions from missing higher orders which
can be done using the residual cutoff dependence as an indicator.
In Fig. \ref{fig:cutoff-selfcon} we display the cutoff dependence of the self-consistent gaps and mean-field shifts obtained within HFB 
($\Delta_1$ and $U_1$), at second order ($\Delta_2$ and $U_2$), and at third order ($\Delta_3$ and $U_3$), for different coupling 
strengths. 

Let us first look at weak coupling $(\kF a)^{-1}=-5$ (leftmost panels), where perturbation theory is expected to work. Indeed, we see 
that with increasing order of perturbation theory, both $\Delta$ and $U$ develop a plateau, i.e., they become approximately cutoff 
independent, in the range $1.5\lesssim\Lambda/\kF \lesssim 3$. If perturbation theory converges to the correct result, the converged 
result must of course be cutoff independent. Hence, the existence of the plateau can be taken as an indication that the perturbation 
expansion up to third order has essentially converged in that range of cutoffs. The noise that is visible, especially at larger 
cutoffs, is due to the amplification of the third-order contribution via the self-consistent procedure as explained in the preceding 
subsection.

When further lowering the cutoff below $\Lambda \lesssim 1.5 \kF$, however, we see that the gap increases again while the mean-field 
gets more attractive, although in this region the perturbation expansion has clearly converged (second- and third-order results lie on 
top of each other). This indicates that there are other missing pieces than higher orders in perturbation theory, such as ``induced'' 
three- and higher-body interactions \cite{Bogner2010} (see Sec. \ref{sect:summary_outlook}).

Let us now look at the more strongly interacting cases $(\kF a)^{-1} = -2$ and $-1$. In these cases, the perturbation expansion does 
not converge so well: the third-order is negligible only in the region where we expect to have corrections from the induced three-body 
interaction. One can still see that a plateau is building up in $U$, although it is limited to a much smaller cutoff interval around 
$\Lambda\approx 2\kF$. In $\Delta$, there is no plateau but only a minimum that can perhaps be understood as the best compromise 
between missing perturbative corrections and missing many-body forces. Nevertheless, the overall variation of $\Delta_3$ in the 
interval $1.5\lesssim \Lambda/\kF \lesssim 3$ is much smaller than the one of the HFB gap $\Delta_1$.

Finally, the rightmost panels show our results in the unitary limit $(\kF a)^{-1}=0$. Although the slope of $U$ as a function of the cutoff gets smaller with increasing order of perturbation theory, there is no convergence, neither in $\Delta$ nor in $U$. This is unfortunate because most experimental studies have focused on the unitary limit. However, if one is mostly interested in cold atoms as simulator of neutron matter, this is less of a problem, because neutron matter is never in such a strongly coupled regime~\cite{Gezerlis2008,Ramanan2013,Calvanese2018,Ramanan2021}.

\begin{figure}[h!]
    \begin{center}
        \includegraphics{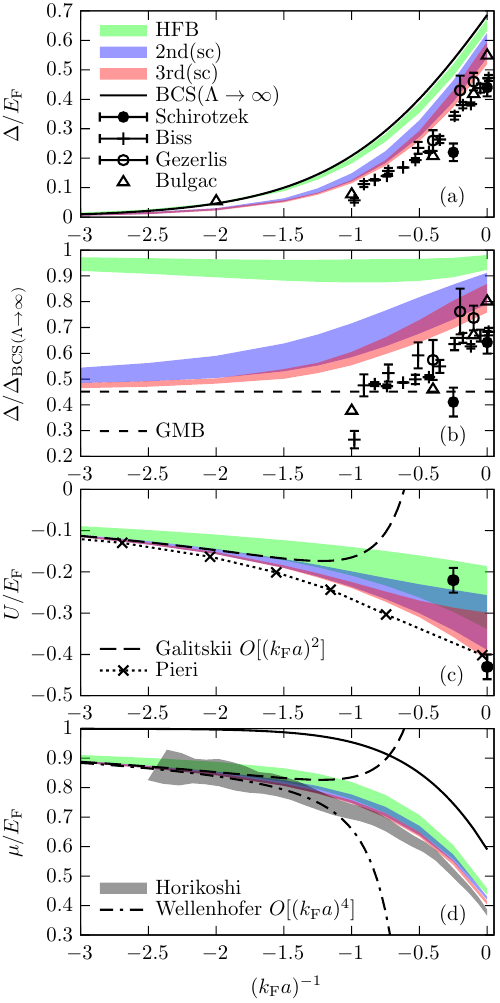}
    \end{center}
    \caption{(a), (b) Gap, (c) mean field shift, and (d) chemical potential as functions of the parameter $1/(\kF a)$, computed in HFB (green) and with self-consistent corrections up to second (blue) and third order (red). For comparison, we show also experimental results by Schirotzek et al.~\cite{Schirotzek2008}, Biss et al. \cite{Biss2022} and Horikoshi et al.~\cite{Horikoshi2017}, and QMC results by Gezerlis and Carlson~\cite{Gezerlis2008} and Bulgac et al.~\cite{Bulgac2008}. The black solid lines show the standard BCS results. Panel (b) shows the gap divided by this BCS gap and the GMB suppression by a factor of $(4e)^{-1/3}$~\cite{Gorkov1961}. Panel (c) also shows the diagrammatic results by Pieri and Pisani, computed as in Ref.~\cite{Pieri2004} but at $T=0$, and the Galitskii result~\cite{Galitskii1958,FetterWalecka}. Panel (d) shows in addition the result of the $\kF a$ expansion by Wellenhofer et al.~\cite{Wellenhofer2021}.}
    \label{fig:selfcon-with-mu}
\end{figure}
\subsection{Comparison with literature and discussion}
We finally show in Fig.~\ref{fig:selfcon-with-mu} our results for the gap, the Hartree term and the chemical potential as functions of $(\kF a)^{-1}$, and compare them with those in the literature. The black solid line, where visible, is the standard BCS result \cite{Calvanese2018} which corresponds in our formalism to the HFB result in the limit $\Lambda \to \infty$. The green bands represent the HFB results, the blue bands include second-order and the red bands up to third-order corrections (within the self-consistent scheme). The bands are obtained by varying the cutoff $\Lambda$ between 
$1.5\kF$ and $2.5\kF$, since this is the range where a plateau (cutoff-independence) starts to build up in the case of convergence as discussed in Sec.~\ref{sec:cutoff-dependence}. Hence, the width of the bands is a measure of the remaining cutoff dependence and represents an estimate of the order of magnitude of missing corrections. However, since the range $1.5 - 2.5\kF$ is somewhat arbitrary, we do not pretend that the bands represent the true error bands of our results.

Let us now discuss each panel in detail. The panels (a) and (b) in Fig.~\ref{fig:selfcon-with-mu} present the gap scaled with $\EF$ and $\Delta_{\text{BCS}}$, respectively.
The HFB gap, which is lower than the BCS result, is further reduced by the inclusion of the second- and third-order self-consistent corrections.
The importance of these corrections is evident in panel (b), especially in the weak-coupling limit, where it is well known that the gap is suppressed by a factor of $(4e)^{-1/3} \approx 0.45$ which is the GMB limit \cite{Gorkov1961}.
As discussed in Sec. \ref{sec:illustration}, this limit is reached only when perturbative corrections of the gap are included \emph{self-consistently}, the second-order corrections achieving almost 50$\%$ reduction in the weak-coupling regime, while the third order adds a smaller reduction for all values of $(\kF a)^{-1}$.
The cutoff dependence seen by the width of the shaded band decreases, especially in the weak coupling limit, when the third-order corrections are included.
However, it increases as one approaches the unitary limit, where the system is strongly correlated and requires higher-order corrections.
Nevertheless, in the unitary regime, within the large uncertainty due to the residual cutoff dependence, our results are compatible with the quantum Monte-Carlo (QMC) results of Gezerlis et al.~\cite{Gezerlis2008} and Bulgac et al.~\cite{Bulgac2008}, as well as the experimental results of Schirotzek et al.~\cite{Schirotzek2008} and Biss et al.~\cite{Biss2022}.
However, away from unitarity, when the system is still strongly coupled, i.e., $(\kF a)^{-1} \sim -1$, the QMC gaps are systematically smaller than our results including third order.
The fact that the experimental gaps are somewhat smaller than ours does not necessarily mean that they are incompatible, as the experiments were done at finite temperature. We will come back to this point below.

The Hartree shift scaled with respect to $\EF$ is shown in panel (c).
As already seen in Sec.~\ref{sec:cutoff-dependence}, the second- and third-order corrections practically eliminate the cutoff dependence in the weak-coupling regime.
However, as in the case of the gap, this is no longer true at strong coupling: the widths of the bands for HFB, second-order, and third-order results become
comparable as $(\kF a)^{-1} \to 0$, indicating the growing size of missing corrections as the system becomes strongly correlated.
Once more, we see that our results, although not converged, are compatible with the experimental result of Schirotzek et al.~\cite{Schirotzek2008} in the unitary limit, but not at $(\kF a)^{-1} = -0.25$.
The long-dashed line presents Galitskii's result 
$U/\EF = \frac{4}{3\pi} \kF a + \frac{4(11-2\ln 2)}{15\pi^2} (\kF a)^2$~\cite{Galitskii1958,FetterWalecka} which is valid in the weak 
coupling region up to order $(\kF a)^2$.
In our calculation, due to the scaling of the cutoff with $\kF$, this is achieved already at 
the HFB level, whereas within the usual regularization scheme ($\Lambda\to\infty$), a resummation of 
ladder diagrams is required~\cite{Perali2002,Pieri2004}.
The dotted line with crosses represents such a resummed Hartree shift computed by Pieri and Pisani as explained in 
reference~\cite{Pieri2004} but at $T=0$.
We note that it is more attractive than ours, but at unitarity the agreement is again reasonable.

Finally, we discuss the chemical potential scaled with $\EF$ in panel (d). The black solid line is the BCS result, while the colored shaded bands are again the HFB and higher-order results.
Surprisingly, the perturbative corrections seem to converge for the entire range of $(\kF a)^{-1}$ considered.
Further, these corrections are essential to reduce the cutoff dependence. 
For $(\kF a)^{-1} < -2$, the perturbative results including third order agree with the dashed-dotted curve, which is the result 
from the $\kF a$ expansion up to fourth order by Wellenhofer et al.~\cite{Wellenhofer2021}.
It is seen that our results lie within the error bands of the experimental results of Horikoshi et al.~\cite{Horikoshi2017} for $(\kF a)^{-1} \lesssim -0.7$ and they stay close to the experimental values of $\mu/\EF$ up to the unitary regime.
The residual differences could arise from the assumptions made about the density being unaffected by the energy- and momentum dependence of the second- and third-order self-energies, as well as the neglected higher-order corrections. It should be noticed that $\mu$ depends on both $\Delta$ and $U$, which both have a strong cutoff dependence near unitarity. Qualitatively, it is easy to understand that their cutoff dependences act in opposite directions: the increase of $U$ with the cutoff tends to also increase $\mu$, while the increase of $\Delta$ with the cutoff tends to decrease $\mu$. However, it is not clear why these two effects cancel almost exactly, as one can see from the surprisingly weak cutoff dependence of $\mu$.

\begin{figure}
    \begin{center}
        \includegraphics{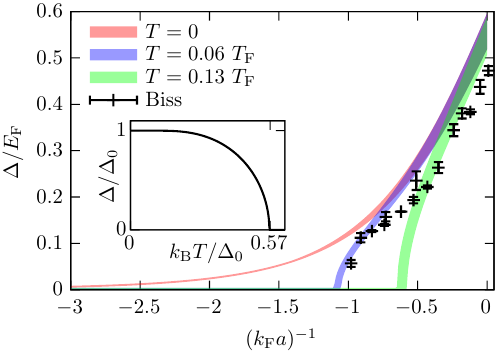}
    \end{center}
    \caption{Rough estimate of how the third-order corrected gap [red band, same as in Fig.~\ref{fig:selfcon-with-mu} (a)] would change at finite temperatures $T/\TF=0.06$ (blue band) and $T/\TF=0.13$ (green band) for a more meaningful comparison with the experiment of Biss et al.~\cite{Biss2022} (data).
    The inset shows the assumed temperature dependence of the gap [with $\Delta_0 = \Delta(T=0)$], according to the weak-coupling BCS theory \cite{Schrieffer,FetterWalecka,Tinkham}.}
    \label{fig:finite-T}
\end{figure}
Let us come back to the comparison with the gaps measured in the experiment by Biss et al.~\cite{Biss2022}.
As stated in Ref.~\cite{Biss2022}, the temperature was $T \approx 0.13~\TF$ at unitarity ($\TF = \EF/\kB$, where $\kB$ is the Boltzmann constant), and when the interaction strength was slowly changed to other values, the entropy per particle stayed constant at $S/N \approx 0.3\, \kB$.
Notice that in the superfluid phase, the entropy as a function of temperature is reduced compared to the one of a normal Fermi gas, $S/(N\kB) = (\pi^2/2) T/\TF$ (see, e.g., Ref.~\cite{Tinkham} or Fig.~3B of Ref.~\cite{Ku2012}).
Therefore, when one starts from the unitary gas in the superfluid phase and then adiabatically reduces the interaction strength towards the weakly interacting BCS regime, thus lowering the critical temperature $T_c$, the temperature of the gas decreases until it reaches $T/\TF = (2/\pi^2) S/(N\kB)$ at the phase transition.
For the parameters of the experiment mentioned above, this suggests that at the weakest couplings the temperature could have been close to $0.06~T_F$.
Unfortunately, our formalism is so far limited to $T=0$, so that we cannot determine either the temperature as a function of the entropy or the temperature dependence of the gap.
To nevertheless estimate at least qualitatively whether temperatures between $0.06$ and $0.13~T_F$ are sufficient to explain the discrepancy between our third-order corrected results and the experimental data, we make the simplifying assumption that the temperature dependence of the gap is the same as in weak-coupling~\cite{FetterWalecka,Schrieffer,Tinkham}, shown in the inset of Fig.~\ref{fig:finite-T}.
In this way we find the blue and green bands shown in Fig.~\ref{fig:finite-T} for $T/\TF=0.06$ and $0.13$, respectively.
When decreasing the interaction strength from unitarity, we expect that one should gradually pass from the green to the blue band as the temperature decreases from $0.13$ to $0.06~\TF$.
In this way, one could interpret the drop of the experimental gap near $(\kF a)^{-1}=-1$ as a sign that at this point the temperature was already quite close to $T_c$. In conclusion, we think that there is no obvious contradiction between our results and the experimental data.

\section{Summary and outlook}
\label{sect:summary_outlook}

In this work, we extend our previous study of Fermi gases with low-momentum effective
interactions \cite{Urban2021} to compute the pairing gap and the Hartree shift. To that end,
we explore the diagrammatic perturbation theory approach using the 
Nambu-Gorkov propagators to obtain corrections to the normal and anomalous self-energy.
The low-momentum effective interactions that have been used depend on the cutoff
$\Lambda$. Since this cutoff is unphysical, any
residual cutoff dependence in the results gives an estimation of the missing corrections, 
which could be from higher orders or from induced higher-body forces, such as three-body force 
\cite{Palaniappan2025}.

In the weak coupling limit, while na\"ive use of perturbation theory works for the Hartree shift if the cutoff is scaled with $\kF$, this formalism fails to reproduce the well-known GMB result for the gap.
However, this limit is recovered provided the gap and the Hartree shift are computed self-consistently in perturbation theory.
In addition, for $(\kF a)^{-1} \lesssim -1$, the perturbation expansion converges in some range of scaled cutoffs around $\Lambda/\kF \sim 2$, and at the same time some cutoff independence within this cutoff range emerges for both the gap and the Hartree shift.
As the system becomes strongly correlated, i.e., as $(\kF a)^{-1} \to 0$, the perturbation theory does not converge well, even at small cutoffs, and hence there is no independence with respect to $\Lambda/\kF$ even when self-consistent corrections are included.
As already mentioned in \cite{Urban2021}, when approaching the BEC side of the BCS-BEC crossover, it might be inefficient to build the perturbation expansion only out of the fermionic excitations, and it could be important to include the Bogoliubov-Anderson mode.

For $\Lambda/\kF < 2$, induced higher-body interactions may start to play a role. 
In fact, although the scattering phase shifts of our effective two-body interaction are by construction independent of the cutoff at low momenta, interactions with different cutoffs are not equivalent at the three- and higher-body level.
This is why reducing the cutoff induces higher-body interactions, even if initially (i.e., for $\Lambda\to\infty$) there is only a two-body interaction \cite{Bogner2010}. 
As we have recently seen in the case of dilute neutron matter \cite{Palaniappan2025}, the three-body interaction induced by the similarity renormalization group evolution of an attractive $s$-wave two-body interaction is repulsive. 
We have shown that once it is included in the calculation of the neutron-matter equation of state, the region of approximate cutoff independence (``plateau'') extends to lower values of $\Lambda/k_F$ (cf. Fig. 8 of \cite{Palaniappan2025}).
Since the induced three-body interaction is repulsive, we expect that it would in the present case make the Hartree shift less attractive and reduce the gap, which is qualitatively what would be needed to reduce also the cutoff dependence of these quantities at $\Lambda/\kF < 2$.
Investigations in this direction are in progress.

In this work we make the additional simplifying assumption that the density is unaffected by the momentum and energy dependence of the second- and third-order self-energy corrections. 
While this approximation simplifies the numerical implementation, it is not guaranteed to work well in the strongly correlated regime.
However, it is worth emphasizing that our results for the gap, the Hartree shift and the chemical potential, when compared against experiments and QMC results, fare reasonably well, in some cases being within the error band of the results in the literature.
Globally, our results for the gap are slightly higher than the QMC results, but they might be compatible with the experimental gaps if one takes into account that the latter were obtained at finite temperature.
Clearly, in order to compare with experiment, it is necessary to extend our formalism to finite temperature.
Looking at Fig.~\ref{fig:selfcon-with-mu}(c), one sees that the existing results for the Hartree shift are quite sparse and not really conclusive.
Additional experimental results, especially at intermediate couplings, would help.

As a next step, it would be of interest to study within this framework the problem of neutron matter, a system that is closely related to the atomic gases examined in this work, but also of relevance to the inner crust of neutron stars.
While the neutron-neutron scattering length is very large, the effective range of the neutron-neutron interaction becomes important with increasing density, and the unitary limit is never realized.
Instead, the ratio $\Delta/\EF$ does not exceed 0.3 \cite{Gezerlis2008}, and in this sense, neutron matter corresponds more to a Fermi gas with $(\kF a)^{-1} < -0.5$. Therefore, one may expect the framework developed in this work to give meaningful results for neutron matter, and we plan to extend our
recent studies \cite{Palaniappan2023,Palaniappan2025}, that considered only corrections to the equation of state, in this direction.
Such a calculation of the pairing gap would be
probably much more reliable than our previous screening
calculations \cite{Ramanan2018,Urban2020} that relied on phenomenological energy-density 
functionals for the Hartree-Fock field and also on some other uncontrolled
approximations such as neglecting the energy dependence of the self-energy.
\acknowledgments
We thank P. Pieri and L. Pisani for useful comments and for sending us their results shown in Fig.~\ref{fig:selfcon-with-mu}(c), and H. Tajima for sending us the data of Ref.~\cite{Horikoshi2017}. This work was supported by CEFIPRA (Collaborative Research Project No. 6304-4). 
\section*{Data availability}
The numerical results shown in the figures are openly available \cite{Urban2026}.

\appendix
\allowdisplaybreaks
\section{Feynman rules} \label{app:feynmanrules}
Here we give the detailed Feynman rules needed to compute the diagrams shown in this paper:
\begin{enumerate}
\item As usual with
  Fermions, we draw the matrix Green's function as a line with an
  arrow, see \Fig{fig:feynmanrules}(a), corresponding to a factor $G^{(0)}_{\alpha\beta}(k,\omega)$.
\item \label{feynmanrule_vertices} 
  The interaction vertices generated by the two-body interaction term $\Hop_{\interaction}$ are diagonal in the Nambu indices. 
  If one of the in- and outgoing Fermion lines has Nambu index 1, the other line must have Nambu index 2 (and vice versa), as shown in
    \Fig{fig:feynmanrules}(b).
  Each interaction is associated with a factor of $-V\big(\tfrac{\kv+\pv'}{2},\tfrac{\kv'+\pv}{2}\big)$, where the negative sign comes from Eq.~\eqref{HintNambu}.
\item The interaction vertices generated by the one-body interaction term $-\Hop_{\meanfield}$ are represented by a cross as shown in \Fig{fig:feynmanrules}(c), and they are associated with a factor of $-H_{\meanfield,\alpha\beta}(k)$.
\item Each vertex conserves energy and momentum, the remaining $N_i$ unconstrained energies and momenta are integrated over with a factor $i^{N_i}/(2\pi)^{4N_i}$,
  unconstrained Nambu indices are summed over, and each closed Fermion loop gives rise to a factor $(-1)$.
  \item If a loop contains only one propagator
    $G^{(0)}_{11}$ or $G^{(0)}_{22}$, it is necessary to multiply it by a factor $e^{i\eta\omega}$ or $e^{-i\eta\omega}$, respectively (with $\eta\to 0^+$) \cite{Schrieffer}, such that the result of the integration over $\omega$ is $\pm 2\pi i v_k^2$, whereas for $G^{(0)}_{12}$ and $G^{(0)}_{21}$ the result of the $\omega$ integration is $-2\pi i u_k v_k$, independently of such a factor.
\end{enumerate}

\section{Symmetry properties of the self-energy}
\label{app:symmetry-properties}

Notice that the matrix elements of $G$ in Eq.~\eqref{GwithAB} have the symmetry properties $G_{12}(p,\omega)=G_{21}(p,\omega)=G_{21}(p,-\omega)$ and $G_{22}(p,\omega) = -G_{11}(p,-\omega)$.
The mean-field term in Eq.~\eqref{eq:Hmeanfield} has similar properties, namely: $H_{\meanfield,12} = H_{\meanfield,21}$ and $H_{\meanfield,22} = -H_{\meanfield,11}$.
Suppose we have calculated a diagram for $\Sigma_{\alpha\beta}(p,\omega)$. 
Then, we can obtain the corresponding $\Sigma_{\bar{\alpha}\bar{\beta}}(p,-\omega)$ by relabeling all internal indices $\alpha_i$ and $\beta_i$ and frequencies $\omega_i$ into $\bar{\alpha}_i$, $\bar{\beta}_i$, and $-\omega_i$, respectively, and then using the symmetry properties of $G$ (and of $H_{\meanfield}$, if it appears in the diagram). 
This gives $\Sigma_{\bar{\alpha}\bar{\beta}}(p,-\omega) = (-1)^{N_d}\Sigma_{\alpha\beta}(p,\omega)$, where $N_d$ denotes the number of diagonal elements of internal propagators ($G_{11}$, $G_{22}$) and mean field terms ($H_{\meanfield,11}$, $H_{\meanfield,22}$).
It is easy to see that $N_d$ is always odd for the normal self-energy ($\alpha=\beta$), leading to Eq.~\eqref{eq:selfenergy-symmetry1a}, while it is even for the anomalous one ($\alpha\ne \beta$), giving Eq.~\eqref{eq:selfenergy-symmetry1b}.

The other symmetry, Eq.~\eqref{eq:selfenergy-symmetry2} is even easier to see. Namely, if one has a diagram for $\Sigma_{12}$, by inverting the directions of all external and internal lines, one gets a diagram for $\Sigma_{21}$. Then in each interaction, incoming lines become outgoing lines and vice versa. However, this does not change its value since $V(\frac{\kv+\pv'}{2},\frac{\kv'+\pv}{2})=V(\frac{\kv'+\pv}{2},\frac{\kv+\pv'}{2})$. As a consequence, when summing all diagrams of a given order (such that for each diagram the corresponding one generated by inverting the directions of the lines is included, too) one obtains Eq.~\eqref{eq:selfenergy-symmetry2}.

\section{Third order diagrams} \label{app:3rdorder}
We explain below the implementation of the Feynman rules for the third order diagrams. 
The diagrams (3.1a)-(3.1d) of the first group have in common the energy integral
\begin{multline}
I^{(3.1)}_{\alpha_1\beta_1\dots\alpha_5\beta_5}(k_1,\dots,k_4) \\
  =-i\int \frac{d{\omega_1}}{2\pi}G_{\alpha_1\beta_1}(k_1{,\omega_1})G_{\alpha_5\beta_5}(k_1{,\omega_1})
  \\
  \times I^{(2)}_{\alpha_2\beta_2\alpha_3\beta_3\alpha_4\beta_4}
    ({\omega_1},k_2,k_3,k_4)
    \label{integral31}
\end{multline}
which we shall denote $I^{(3.1)}_{12345}$.
In the same way, let us denote $A_{\alpha_1\beta_1}(k_1) = A_1$, $E_{k_1,k_2,k_3} = E_{123}$ etc., and we use $\kv_5 = \kv_1$ such that, e.g., $A_5 = A_{\alpha_5\beta_5}(k_1)$.
Furthermore, we introduce the abbreviation $E^- = E-i\eta$.
In this notation, e.g., Eq.~\eqref{integral2} can be compactly rewritten as
\begin{equation}
I^{(2)}_{123}({\omega}) = -\frac{A_1A_2B_3}{{\omega}-E^{{-}}_{123}}
  -\frac{B_1B_2A_3}{{\omega}+E^{{-}}_{123}}\,.
\end{equation}
Closing as usual the contour in the complex plane, one can easily perform the ${\omega_1}$ integration in Eq.~\eqref{integral31}, with the result
\begin{multline}
I^{(3.1)}_{12345} = \frac{B_1B_5A_2A_3B_4-A_1A_5B_2B_3A_4}{E_{1234}^2}\\
  + \frac{(A_1B_5+B_1A_5)(B_2B_3A_4-A_2A_3B_4)}{2E_1E_{1234}}\,.
\end{multline}
Then, each individual diagram $(3.1i)$ ($i = $ a\dots d) is obtained by multiplying this expression with the respective combinations of vertices $V^{(3.1i)}_{\text{A}}$, $V^{(3.1i)}_{\text{B}}$, $V^{(3.1i)}_{\text{C}}$ and the sign corresponding to the number of closed Fermion loops $F^{(3.1i)}$, integrating over $\kv_1$, $\kv_2$, and $\kv_3$, noting that $\kv_4 = \kv_2+\kv_3-\kv_1$ (and $\kv_5 = \kv_1$ as already mentioned), and summing over the internal indices:
\begin{multline}
  \Sigma^{(3.1i)}_{\alpha\beta} = \int\frac{d^3k_1}{(2\pi)^3}
  \int\frac{d^3k_2}{(2\pi)^3} \int\frac{d^3k_3}{(2\pi)^3} (-1)^{F^{(3.1i)}}\\
  \times\sum_{\alpha_1\beta_1\dots\alpha_5\beta_5} 
  V^{(3.1i)}_{\text{A}} 
  V^{(3.1i)}_{\text{B}} 
  V^{(3.1i)}_{\text{C}} 
  I^{(3.1)}_{12345}\,,
\end{multline}
with
\begin{align}
  F^{(3.1i)} &= 
    \begin{cases}
      2\,,& {i = \mathrm{a},}\\
      1\,,& {i = \mathrm{b,c},}\\
      0\,,& {i = \mathrm{d},}
    \end{cases}\\
  V^{(3.1i)}_{\text{A}} &= 
    \begin{cases}
      V(\tfrac{\pv+\kv_1}{2},\tfrac{\pv+\kv_1}{2})
      \delta_{\alpha\beta}
      \delta_{\bar{\alpha}\alpha_1}
      \delta_{\bar{\alpha}\beta_5}\,,
    &{i = \mathrm{a,b},}\\
      V(\pv,\kv_1)
      \delta_{\bar{\alpha}
      \beta}\delta_{\alpha\alpha_1}
      \delta_{\bar{\alpha}\beta_5}\,,
    &{i = \mathrm{c,d},}
    \end{cases}\\
  V^{(3.1{i})}_{\text{B}} &=
    \begin{cases}
      V(\tfrac{\kv_1+\kv_3}{2},\tfrac{\kv_2+\kv_4}{2})
      \delta_{\beta_2\alpha_5}
      \delta_{\bar{\beta}_2\beta_3}
      \delta_{\bar{\beta}_2\alpha_4}\,,
    &{i = \mathrm{a,c},}\\
      V(\tfrac{\kv_1+\kv_2}{2},\tfrac{\kv_3+\kv_4}{2})
      \delta_{\beta_2\alpha_4}
      \delta_{\bar{\beta}_2\beta_3}
      \delta_{\bar{\beta}_2\alpha_5}\,,
    &{i = \mathrm{b,d}.}
    \end{cases}\\
  V^{(3.1i)}_{\text{C}} &=
    V(\tfrac{\kv_1+\kv_3}{2},\tfrac{\kv_2+\kv_4}{2})
    \delta_{\alpha_2\beta_1}
    \delta_{\bar{\alpha}_2\alpha_3}
    \delta_{\bar{\alpha}_2\beta_4}\,.
\end{align}
Notice that the Kronecker deltas determine all internal indices except two, say, $\alpha_2$ and $\beta_2$.

Let us now compute the diagrams of the second group, (3.2a) and (3.2b). It is convenient to define the ``particle-particle like'' convolution of two propagators (remember that, depending on the combination of Nambu indices, it describes particle-particle, hole-hole, particle-hole, or anomalous channels):
\begin{align}
  \Pi_{12}({\omega}) &= -i \int\frac{d{\omega_1}}{(2\pi)} G_1({\omega_1})G_2({\omega-\omega_1})\nonumber\\
    &= -\frac{A_1A_2}{{\omega}-E^{{-}}_{12}}+\frac{B_1B_2}{{\omega}+E^{{-}}_{12}}\,,
\end{align}
where we have again used the same short-hand notations as before and $G_1({\omega}) = G_{\alpha_1\beta_1}(k_1,{\omega})$ etc.
With this ingredient, the energy integral appearing in the diagrams (3.2a) and (3.2b) reads
\begin{align}
  I^{(3.2)}_{12345}({\omega}) =& -i\int \frac{d{\omega'}}{(2\pi)^3}G_1({\omega'}-{\omega})
    \Pi_{23}({\omega'})\Pi_{45}({\omega'})\nonumber\\
    =& -\frac{B_1A_2A_3B_4B_5}{E_{2345}({\omega}-E^{{-}}_{123})}
      -\frac{B_1B_2B_3A_4A_5}{E_{2345}({\omega}-E^{{-}}_{145})}\nonumber\\
      &-\frac{A_1B_2B_3A_4A_5}{E_{2345}({\omega}+E^{{-}}_{123})}
      -\frac{A_1A_2A_3B_4B_5}{E_{2345}({\omega}+E^{{-}}_{145})}\nonumber\\
      &+\frac{B_1A_2A_3A_4A_5}{({\omega}-E^{{-}}_{123})({\omega}-E^{{-}}_{145})}\nonumber\\
      &-\frac{A_1B_2B_3B_4B_5}{({\omega}+E^{{-}}_{123})({\omega}+E^{{-}}_{145})}\,.
      \label{integral32}
\end{align}
The five momenta are not independent but they have to satisfy $\pv+\kv_1 = \kv_2+\kv_3 = \kv_4+\kv_5$, i.e., we can choose, e.g., $\kv_1$, $\kv_2$, and $\kv_4$ as integration variables.
Denoting the vertices in each diagram (3.2$i$) by $V^{(3.2i)}_{\text{A}}$, $V^{(3.2i)}_{\text{B}}$, and $V^{(3.2i)}_{\text{C}}$, we can write the contribution of diagram (3.2$i$) as
\begin{multline}
  \Sigma^{(3.2i)}_{\alpha\beta}({\omega}) = \int\frac{d^3k_1}{(2\pi)^3}
  \int\frac{d^3k_2}{(2\pi)^3} \int\frac{d^3k_4}{(2\pi)^3} (-1)^{F^{(3.2i)}}\\
  {\times}\sum_{\alpha_1\beta_1\dots\alpha_5\beta_5} 
  V^{(3.2i)}_{\text{A}} 
  V^{(3.2i)}_{\text{B}}
  V^{(3.2i)}_{\text{C}} 
  I^{(3.2)}_{12345}({\omega})\,,
  \label{diagram32i}
\end{multline}
with
\begin{align}
  F^{(3.2i)} =&
  \begin{cases}
        1\,,& {i = \mathrm{a},}\\
        0\,,& {i = \mathrm{b},}
    \end{cases}\\
    V^{(3.2i)}_{\text{A}} =& V(\tfrac{\pv+\kv_2}{2},\tfrac{\kv_1+\kv_3}{2})
    \delta_{\beta\beta_3}\delta_{\bar{\beta}\alpha_1}\delta_{\bar{\beta}\beta_2}\,,\\
  V^{(3.2i)}_{\text{B}} =& 
    \begin{cases}
      V(\tfrac{\kv_2+\kv_5}{2},\tfrac{\kv_3+\kv_4}{2})
      \delta_{\alpha_2\beta_4}
      \delta_{\bar{\alpha}_2\alpha_3}
      \delta_{\bar{\alpha}_2\beta_5}\,,
    &{i = \mathrm{a},} \\
      V(\tfrac{\kv_2+\kv_4}{2},\tfrac{\kv_3+\kv_5}{2})
      \delta_{\alpha_2\beta_5}
      \delta_{\bar{\alpha}_2\alpha_3}
      \delta_{\bar{\alpha}_2\beta_4}\,,
    &{i = \mathrm{b},} \\
    \end{cases} \\
  V^{(3.2i)}_{\text{C}} =&
    V(\tfrac{\pv+\kv_4}{2},\tfrac{\kv_1+\kv_5}{2})
    \delta_{\alpha\alpha_5}\delta_{\bar{\alpha}\beta_1}\delta_{\bar{\alpha}\alpha_4}\,.
\end{align}
We see that the Kronecker deltas determine all internal indices except one, e.g., $\alpha_2$.

Finally, we turn to the diagrams of the third group, (3.3a-h). Here, we need the ``particle-hole like'' propagator (again, this name is a bit misleading because it contains also particle-particle, hole-hole, and anomalous contributions):
\begin{align}
  \chi_{12}({\omega}) &= -i \int\frac{d{\omega_1}}{(2\pi)} G_1({\omega_1})G_2({\omega}+{\omega_1})\nonumber\\
    &= -\frac{B_1A_2}{{\omega}-E^{{-}}_{12}}-\frac{A_1B_2}{{\omega}+E^{{-}}_{12}}\,.
\end{align}
The common energy integral in the diagrams (3.3a-h) reads
\begin{align}
  I^{(3.3)}_{12345}({\omega}) =& -i\int \frac{d{\omega'}}{(2\pi)^3}G_1({\omega}-{\omega'})
    \chi_{23}({\omega'})\chi_{45}({\omega'})\nonumber\\
    =& \frac{A_1B_2A_3A_4B_5}{E_{2345}({\omega}-E^{{-}}_{123})}
      +\frac{A_1A_2B_3B_4A_5}{E_{2345}({\omega}-E^{{-}}_{145})}\nonumber\\
      &+\frac{B_1A_2B_3B_4A_5}{E_{2345}({\omega}+E^{{-}}_{123})}
      +\frac{B_1B_2A_3A_4B_5}{E_{2345}({\omega}+E^{{-}}_{145})}\nonumber\\
      &-\frac{A_1B_2A_3B_4A_5}{({\omega}-E^{{-}}_{123})({\omega}-E^{{-}}_{145})}\nonumber\\
      &+\frac{B_1A_2B_3A_4B_5}{({\omega}+E^{{-}}_{123})({\omega}+E^{{-}}_{145})}\,.
      \label{integral33}
\end{align}
We see the same energy denominators as in Eq.~\eqref{integral32}, but the $A$ and $B$ (i.e., $u$ and $v$) factors in the numerators appear in different combinations.
Again, we can choose three out of the five momenta $\kv_1\dots \kv_5$ as integration variables, the two remaining momenta are then determined by the relationships $\pv-\kv_1 = \kv_3-\kv_2 = \kv_5-\kv_4$.
Analogously to Eq.~\eqref{diagram32i}, we write
\begin{multline}
  \Sigma^{(3.3i)}_{\alpha\beta}({\omega}) = \int\frac{d^3k_1}{(2\pi)^3}
  \int\frac{d^3k_2}{(2\pi)^3} \int\frac{d^3k_4}{(2\pi)^3} (-1)^{F^{({3.}3i)}}\\
  {\times}\sum_{\alpha_1\beta_1\dots\alpha_5\beta_5} 
  V^{(3.3i)}_{\text{{A}}} 
  V^{(3.3i)}_{\text{{B}}}
  V^{(3.3i)}_{\text{{C}}} 
  I^{(3.3)}_{12345}({\omega})\,,
  \label{diagram33i}
\end{multline}
with
\begin{align}
  F^{(3.3i)} =&
    \begin{cases}
      1\,,& {i = \mathrm{a,d,f,g},}\\
      0\,,& {i = \mathrm{b,c,e},}\\
      2\,,& {i = \mathrm{h},}
    \end{cases}\\
    V^{(3.3i)}_{\text{A}} =& 
    \begin{cases}
      V(\tfrac{\pv+\kv_1}{2},\tfrac{\kv_2+\kv_3}{2})
      \delta_{\beta\beta_3}
      \delta_{\bar{\beta}\beta_1}
      \delta_{\bar{\beta}\alpha_2}\,,
    & {i = \mathrm{a,b,c,d},}\\
      V(\tfrac{\pv+\kv_3}{2},\tfrac{\kv_1+\kv_2}{2})
      \delta_{\beta\beta_1}
      \delta_{\bar{\beta}\alpha_2}
      \delta_{\bar{\beta}\beta_3}\,,
    & {i = \mathrm{e,f,g,h},}
  \end{cases}\\
  V^{(3.3i)}_{\text{{B}}} =&
    \mbox{\scalebox{0.95}[1]{$\displaystyle\,\begin{cases}
      V(\tfrac{\kv_2+\kv_3}{2},\tfrac{\kv_4+\kv_5}{2})
      \delta_{\beta_2\alpha_4}
      \delta_{\bar{\beta}_2\alpha_3}
      \delta_{\bar{\beta}_2\beta_5}\,,
    & {i = \mathrm{a,c,e,g},}\\
      V(\tfrac{\kv_2+\kv_4}{2},\tfrac{\kv_3+\kv_5}{2})
      \delta_{\beta_2\alpha_3}
      \delta_{\bar{\beta}_2\alpha_4}
      \delta_{\bar{\beta}_2\beta_5}\,,
    & {i = \mathrm{b,d,f,h},}
    \end{cases}$}}\\    
  V^{(3.3i)}_{\text{{C}}} =&
    \begin{cases}
      V(\tfrac{\pv+\kv_1}{2},\tfrac{\kv_4+\kv_5}{2})
      \delta_{\alpha\alpha_5}
      \delta_{\bar{\alpha}\alpha_1}
      \delta_{\bar{\alpha}\beta_4}\,,
    & {i = \mathrm{a,b,e,f},}\\
      V(\tfrac{\pv+\kv_5}{2},\tfrac{\kv_1+\kv_4}{2})
      \delta_{\alpha\alpha_1}
      \delta_{\bar{\alpha}\alpha_5}
      \delta_{\bar{\alpha}\beta_4}\,,
    & {i = \mathrm{c,d,g,h}.}
    \end{cases}
\end{align}
Again, for given $\alpha$ and $\beta$, there is only one summation
over Nambu indices ($\beta_2$) left.

\bibliography{references}

\begin{thebibliography}{35}%
\makeatletter
\providecommand \@ifxundefined [1]{%
 \@ifx{#1\undefined}
}%
\providecommand \@ifnum [1]{%
 \ifnum #1\expandafter \@firstoftwo
 \else \expandafter \@secondoftwo
 \fi
}%
\providecommand \@ifx [1]{%
 \ifx #1\expandafter \@firstoftwo
 \else \expandafter \@secondoftwo
 \fi
}%
\providecommand \natexlab [1]{#1}%
\providecommand \enquote  [1]{``#1''}%
\providecommand \bibnamefont  [1]{#1}%
\providecommand \bibfnamefont [1]{#1}%
\providecommand \citenamefont [1]{#1}%
\providecommand \href@noop [0]{\@secondoftwo}%
\providecommand \href [0]{\begingroup \@sanitize@url \@href}%
\providecommand \@href[1]{\@@startlink{#1}\@@href}%
\providecommand \@@href[1]{\endgroup#1\@@endlink}%
\providecommand \@sanitize@url [0]{\catcode `\\12\catcode `\$12\catcode
  `\&12\catcode `\#12\catcode `\^12\catcode `\_12\catcode `\%12\relax}%
\providecommand \@@startlink[1]{}%
\providecommand \@@endlink[0]{}%
\providecommand \url  [0]{\begingroup\@sanitize@url \@url }%
\providecommand \@url [1]{\endgroup\@href {#1}{\urlprefix }}%
\providecommand \urlprefix  [0]{URL }%
\providecommand \Eprint [0]{\href }%
\providecommand \doibase [0]{https://doi.org/}%
\providecommand \selectlanguage [0]{\@gobble}%
\providecommand \bibinfo  [0]{\@secondoftwo}%
\providecommand \bibfield  [0]{\@secondoftwo}%
\providecommand \translation [1]{[#1]}%
\providecommand \BibitemOpen [0]{}%
\providecommand \bibitemStop [0]{}%
\providecommand \bibitemNoStop [0]{.\EOS\space}%
\providecommand \EOS [0]{\spacefactor3000\relax}%
\providecommand \BibitemShut  [1]{\csname bibitem#1\endcsname}%
\let\auto@bib@innerbib\@empty
\bibitem [{\citenamefont {Ku}\ \emph {et~al.}(2012)\citenamefont {Ku},
  \citenamefont {Sommer}, \citenamefont {Cheuk},\ and\ \citenamefont
  {Zwierlein}}]{Ku2012}%
  \BibitemOpen
  \bibfield  {author} {\bibinfo {author} {\bibfnamefont {M.~J.~H.}\
  \bibnamefont {Ku}}, \bibinfo {author} {\bibfnamefont {A.~T.}\ \bibnamefont
  {Sommer}}, \bibinfo {author} {\bibfnamefont {L.~W.}\ \bibnamefont {Cheuk}},\
  and\ \bibinfo {author} {\bibfnamefont {M.~W.}\ \bibnamefont {Zwierlein}},\
  }\bibfield  {title} {\bibinfo {title} {{Revealing the Superfluid Lambda
  Transition in the Universal Thermodynamics of a Unitary Fermi Gas}},\ }\href
  {https://doi.org/10.1126/science.1214987} {\bibfield  {journal} {\bibinfo
  {journal} {Science}\ }\textbf {\bibinfo {volume} {335}},\ \bibinfo {pages}
  {563} (\bibinfo {year} {2012})}\BibitemShut {NoStop}%
\bibitem [{\citenamefont {Horikoshi}\ \emph {et~al.}(2017)\citenamefont
  {Horikoshi}, \citenamefont {Koashi}, \citenamefont {Tajima}, \citenamefont
  {Ohashi},\ and\ \citenamefont {Kuwata-Gonokami}}]{Horikoshi2017}%
  \BibitemOpen
  \bibfield  {author} {\bibinfo {author} {\bibfnamefont {M.}~\bibnamefont
  {Horikoshi}}, \bibinfo {author} {\bibfnamefont {M.}~\bibnamefont {Koashi}},
  \bibinfo {author} {\bibfnamefont {H.}~\bibnamefont {Tajima}}, \bibinfo
  {author} {\bibfnamefont {Y.}~\bibnamefont {Ohashi}},\ and\ \bibinfo {author}
  {\bibfnamefont {M.}~\bibnamefont {Kuwata-Gonokami}},\ }\bibfield  {title}
  {\bibinfo {title} {{Ground-State Thermodynamic Quantities of Homogeneous
  Spin-$1/2$ Fermions from the BCS Region to the Unitarity Limit}},\ }\href
  {https://doi.org/10.1103/PhysRevX.7.041004} {\bibfield  {journal} {\bibinfo
  {journal} {Phys. Rev. X}\ }\textbf {\bibinfo {volume} {7}},\ \bibinfo {pages}
  {041004} (\bibinfo {year} {2017})}\BibitemShut {NoStop}%
\bibitem [{\citenamefont {{Calvanese Strinati}}\ \emph
  {et~al.}(2018)\citenamefont {{Calvanese Strinati}}, \citenamefont {{Pieri}},
  \citenamefont {{R\"opke}}, \citenamefont {{Schuck}},\ and\ \citenamefont
  {{Urban}}}]{Calvanese2018}%
  \BibitemOpen
  \bibfield  {author} {\bibinfo {author} {\bibfnamefont {G.}~\bibnamefont
  {{Calvanese Strinati}}}, \bibinfo {author} {\bibfnamefont {P.}~\bibnamefont
  {{Pieri}}}, \bibinfo {author} {\bibfnamefont {G.}~\bibnamefont {{R\"opke}}},
  \bibinfo {author} {\bibfnamefont {P.}~\bibnamefont {{Schuck}}},\ and\
  \bibinfo {author} {\bibfnamefont {M.}~\bibnamefont {{Urban}}},\ }\bibfield
  {title} {\bibinfo {title} {{The BCS-BEC crossover: From ultra-cold Fermi
  gases to nuclear systems}},\ }\href
  {https://doi.org/https://doi.org/10.1016/j.physrep.2018.02.004} {\bibfield
  {journal} {\bibinfo  {journal} {Phys. Rep.}\ }\textbf {\bibinfo {volume}
  {738}},\ \bibinfo {pages} {1} (\bibinfo {year} {2018})}\BibitemShut {NoStop}%
\bibitem [{\citenamefont {Baker}(1999)}]{Baker1999}%
  \BibitemOpen
  \bibfield  {author} {\bibinfo {author} {\bibfnamefont {G.~A.}\ \bibnamefont
  {Baker}},\ }\bibfield  {title} {\bibinfo {title} {{Neutron matter model}},\
  }\href {https://doi.org/10.1103/PhysRevC.60.054311} {\bibfield  {journal}
  {\bibinfo  {journal} {Phys. Rev. C}\ }\textbf {\bibinfo {volume} {60}},\
  \bibinfo {pages} {054311} (\bibinfo {year} {1999})}\BibitemShut {NoStop}%
\bibitem [{\citenamefont {Stewart}\ \emph {et~al.}(2008)\citenamefont
  {Stewart}, \citenamefont {Gaebler},\ and\ \citenamefont {Jin}}]{Stewart2008}%
  \BibitemOpen
  \bibfield  {author} {\bibinfo {author} {\bibfnamefont {J.}~\bibnamefont
  {Stewart}}, \bibinfo {author} {\bibfnamefont {J.}~\bibnamefont {Gaebler}},\
  and\ \bibinfo {author} {\bibfnamefont {D.}~\bibnamefont {Jin}},\ }\bibfield
  {title} {\bibinfo {title} {{Using photoemission spectroscopy to probe a
  strongly interacting Fermi gas}},\ }\href
  {https://doi.org/10.1038/nature07172} {\bibfield  {journal} {\bibinfo
  {journal} {Nature}\ }\textbf {\bibinfo {volume} {454}},\ \bibinfo {pages}
  {744} (\bibinfo {year} {2008})}\BibitemShut {NoStop}%
\bibitem [{\citenamefont {Gor'kov}\ and\ \citenamefont
  {Melik-Barkhudarov}(1961)}]{Gorkov1961}%
  \BibitemOpen
  \bibfield  {author} {\bibinfo {author} {\bibfnamefont {L.~P.}\ \bibnamefont
  {Gor'kov}}\ and\ \bibinfo {author} {\bibfnamefont {T.~K.}\ \bibnamefont
  {Melik-Barkhudarov}},\ }\bibfield  {title} {\bibinfo {title} {{Contribution
  to the theory of superfluidity in an imperfect Fermi gas}},\ }\href@noop {}
  {\bibfield  {journal} {\bibinfo  {journal} {Zh. Exp. Teor. Fiz.}\ }\textbf
  {\bibinfo {volume} {40}},\ \bibinfo {pages} {1452} (\bibinfo {year}
  {1961})}\BibitemShut {NoStop}%
\bibitem [{\citenamefont {Gezerlis}\ and\ \citenamefont
  {Carlson}(2008)}]{Gezerlis2008}%
  \BibitemOpen
  \bibfield  {author} {\bibinfo {author} {\bibfnamefont {A.}~\bibnamefont
  {Gezerlis}}\ and\ \bibinfo {author} {\bibfnamefont {J.}~\bibnamefont
  {Carlson}},\ }\bibfield  {title} {\bibinfo {title} {{Strongly paired
  fermions: Cold atoms and neutron matter}},\ }\href
  {https://doi.org/10.1103/PhysRevC.77.032801} {\bibfield  {journal} {\bibinfo
  {journal} {Phys. Rev. C}\ }\textbf {\bibinfo {volume} {77}},\ \bibinfo
  {pages} {032801} (\bibinfo {year} {2008})}\BibitemShut {NoStop}%
\bibitem [{\citenamefont {Biss}\ \emph {et~al.}(2022)\citenamefont {Biss},
  \citenamefont {Sobirey}, \citenamefont {Luick}, \citenamefont {Bohlen},
  \citenamefont {Kinnunen}, \citenamefont {Bruun}, \citenamefont {Lompe},\ and\
  \citenamefont {Moritz}}]{Biss2022}%
  \BibitemOpen
  \bibfield  {author} {\bibinfo {author} {\bibfnamefont {H.}~\bibnamefont
  {Biss}}, \bibinfo {author} {\bibfnamefont {L.}~\bibnamefont {Sobirey}},
  \bibinfo {author} {\bibfnamefont {N.}~\bibnamefont {Luick}}, \bibinfo
  {author} {\bibfnamefont {M.}~\bibnamefont {Bohlen}}, \bibinfo {author}
  {\bibfnamefont {J.~J.}\ \bibnamefont {Kinnunen}}, \bibinfo {author}
  {\bibfnamefont {G.~M.}\ \bibnamefont {Bruun}}, \bibinfo {author}
  {\bibfnamefont {T.}~\bibnamefont {Lompe}},\ and\ \bibinfo {author}
  {\bibfnamefont {H.}~\bibnamefont {Moritz}},\ }\bibfield  {title} {\bibinfo
  {title} {{Excitation Spectrum and Superfluid Gap of an Ultracold Fermi
  Gas}},\ }\href {https://doi.org/10.1103/PhysRevLett.128.100401} {\bibfield
  {journal} {\bibinfo  {journal} {Phys. Rev. Lett.}\ }\textbf {\bibinfo
  {volume} {128}},\ \bibinfo {pages} {100401} (\bibinfo {year}
  {2022})}\BibitemShut {NoStop}%
\bibitem [{\citenamefont {Pisani}\ \emph {et~al.}(2018)\citenamefont {Pisani},
  \citenamefont {Pieri},\ and\ \citenamefont {Strinati}}]{Pisani2018}%
  \BibitemOpen
  \bibfield  {author} {\bibinfo {author} {\bibfnamefont {L.}~\bibnamefont
  {Pisani}}, \bibinfo {author} {\bibfnamefont {P.}~\bibnamefont {Pieri}},\ and\
  \bibinfo {author} {\bibfnamefont {G.~C.}\ \bibnamefont {Strinati}},\
  }\bibfield  {title} {\bibinfo {title} {{Gap equation with pairing
  correlations beyond the mean-field approximation and its equivalence to a
  Hugenholtz-Pines condition for fermion pairs}},\ }\href
  {https://doi.org/10.1103/PhysRevB.98.104507} {\bibfield  {journal} {\bibinfo
  {journal} {Phys. Rev. B}\ }\textbf {\bibinfo {volume} {98}},\ \bibinfo
  {pages} {104507} (\bibinfo {year} {2018})}\BibitemShut {NoStop}%
\bibitem [{\citenamefont {Urban}\ and\ \citenamefont
  {Ramanan}(2021)}]{Urban2021}%
  \BibitemOpen
  \bibfield  {author} {\bibinfo {author} {\bibfnamefont {M.}~\bibnamefont
  {Urban}}\ and\ \bibinfo {author} {\bibfnamefont {S.}~\bibnamefont
  {Ramanan}},\ }\bibfield  {title} {\bibinfo {title} {{Low-momentum
  interactions for ultracold Fermi gases}},\ }\href
  {https://doi.org/10.1103/PhysRevA.103.063306} {\bibfield  {journal} {\bibinfo
   {journal} {Phys. Rev. A}\ }\textbf {\bibinfo {volume} {103}},\ \bibinfo
  {pages} {063306} (\bibinfo {year} {2021})}\BibitemShut {NoStop}%
\bibitem [{\citenamefont {Ramanan}\ and\ \citenamefont
  {Urban}(2018)}]{Ramanan2018}%
  \BibitemOpen
  \bibfield  {author} {\bibinfo {author} {\bibfnamefont {S.}~\bibnamefont
  {Ramanan}}\ and\ \bibinfo {author} {\bibfnamefont {M.}~\bibnamefont
  {Urban}},\ }\bibfield  {title} {\bibinfo {title} {{Screening and
  antiscreening of the pairing interaction in low-density neutron matter}},\
  }\href {https://doi.org/10.1103/PhysRevC.98.024314} {\bibfield  {journal}
  {\bibinfo  {journal} {Phys. Rev. C}\ }\textbf {\bibinfo {volume} {98}},\
  \bibinfo {pages} {024314} (\bibinfo {year} {2018})}\BibitemShut {NoStop}%
\bibitem [{\citenamefont {Pieri}\ and\ \citenamefont
  {Strinati}(2000)}]{Pieri2000}%
  \BibitemOpen
  \bibfield  {author} {\bibinfo {author} {\bibfnamefont {P.}~\bibnamefont
  {Pieri}}\ and\ \bibinfo {author} {\bibfnamefont {G.~C.}\ \bibnamefont
  {Strinati}},\ }\bibfield  {title} {\bibinfo {title} {{Strong-coupling limit
  in the evolution from BCS superconductivity to Bose-Einstein condensation}},\
  }\href {https://doi.org/10.1103/PhysRevB.61.15370} {\bibfield  {journal}
  {\bibinfo  {journal} {Phys. Rev. B}\ }\textbf {\bibinfo {volume} {61}},\
  \bibinfo {pages} {15370} (\bibinfo {year} {2000})}\BibitemShut {NoStop}%
\bibitem [{\citenamefont {S\'a~de Melo}\ \emph {et~al.}(1993)\citenamefont
  {S\'a~de Melo}, \citenamefont {Randeria},\ and\ \citenamefont
  {Engelbrecht}}]{SadeMelo1993}%
  \BibitemOpen
  \bibfield  {author} {\bibinfo {author} {\bibfnamefont {C.~A.~R.}\
  \bibnamefont {S\'a~de Melo}}, \bibinfo {author} {\bibfnamefont
  {M.}~\bibnamefont {Randeria}},\ and\ \bibinfo {author} {\bibfnamefont
  {J.~R.}\ \bibnamefont {Engelbrecht}},\ }\bibfield  {title} {\bibinfo {title}
  {{Crossover from BCS to Bose superconductivity: Transition temperature and
  time-dependent Ginzburg-Landau theory}},\ }\href
  {https://doi.org/10.1103/PhysRevLett.71.3202} {\bibfield  {journal} {\bibinfo
   {journal} {Phys. Rev. Lett.}\ }\textbf {\bibinfo {volume} {71}},\ \bibinfo
  {pages} {3202} (\bibinfo {year} {1993})}\BibitemShut {NoStop}%
\bibitem [{\citenamefont {Bogner}\ \emph {et~al.}(2005)\citenamefont {Bogner},
  \citenamefont {Schwenk}, \citenamefont {Furnstahl},\ and\ \citenamefont
  {Nogga}}]{Bogner2005}%
  \BibitemOpen
  \bibfield  {author} {\bibinfo {author} {\bibfnamefont {S.~K.}\ \bibnamefont
  {Bogner}}, \bibinfo {author} {\bibfnamefont {A.}~\bibnamefont {Schwenk}},
  \bibinfo {author} {\bibfnamefont {R.~J.}\ \bibnamefont {Furnstahl}},\ and\
  \bibinfo {author} {\bibfnamefont {A.}~\bibnamefont {Nogga}},\ }\bibfield
  {title} {\bibinfo {title} {{Is nuclear matter perturbative with low-momentum
  interactions?}},\ }\href {https://doi.org/10.1016/j.nuclphysa.2005.08.024}
  {\bibfield  {journal} {\bibinfo  {journal} {Nucl. Phys. A}\ }\textbf
  {\bibinfo {volume} {763}},\ \bibinfo {pages} {59} (\bibinfo {year}
  {2005})}\BibitemShut {NoStop}%
\bibitem [{\citenamefont {Schwenk}\ \emph {et~al.}(2003)\citenamefont
  {Schwenk}, \citenamefont {Friman},\ and\ \citenamefont
  {Brown}}]{Schwenk2003}%
  \BibitemOpen
  \bibfield  {author} {\bibinfo {author} {\bibfnamefont {A.}~\bibnamefont
  {Schwenk}}, \bibinfo {author} {\bibfnamefont {B.}~\bibnamefont {Friman}},\
  and\ \bibinfo {author} {\bibfnamefont {G.~E.}\ \bibnamefont {Brown}},\
  }\bibfield  {title} {\bibinfo {title} {{Renormalization group approach to
  neutron matter: quasiparticle interactions, superfluid gaps and the equation
  of state}},\ }\href
  {https://doi.org/https://doi.org/10.1016/S0375-9474(02)01290-3} {\bibfield
  {journal} {\bibinfo  {journal} {Nucl. Phys. A}\ }\textbf {\bibinfo {volume}
  {713}},\ \bibinfo {pages} {191} (\bibinfo {year} {2003})}\BibitemShut
  {NoStop}%
\bibitem [{\citenamefont {Taillat}\ and\ \citenamefont
  {Kurkjian}(2025)}]{Taillat2025}%
  \BibitemOpen
  \bibfield  {author} {\bibinfo {author} {\bibfnamefont {P.-L.}\ \bibnamefont
  {Taillat}}\ and\ \bibinfo {author} {\bibfnamefont {H.}~\bibnamefont
  {Kurkjian}},\ }\href {https://arxiv.org/abs/2511.15938} {\bibinfo {title} {{A
  low-energy effective Hamiltonian for Landau quasiparticles}}} (\bibinfo
  {year} {2025}),\ \Eprint {https://arxiv.org/abs/2511.15938} {arXiv:2511.15938
  [cond-mat.quant-gas]} \BibitemShut {NoStop}%
\bibitem [{\citenamefont {Urban}\ and\ \citenamefont
  {Ramanan}(2020)}]{Urban2020}%
  \BibitemOpen
  \bibfield  {author} {\bibinfo {author} {\bibfnamefont {M.}~\bibnamefont
  {Urban}}\ and\ \bibinfo {author} {\bibfnamefont {S.}~\bibnamefont
  {Ramanan}},\ }\bibfield  {title} {\bibinfo {title} {{Neutron pairing with
  medium polarization beyond the Landau approximation}},\ }\href
  {https://doi.org/10.1103/PhysRevC.101.035803} {\bibfield  {journal} {\bibinfo
   {journal} {Phys. Rev. C}\ }\textbf {\bibinfo {volume} {101}},\ \bibinfo
  {pages} {035803} (\bibinfo {year} {2020})}\BibitemShut {NoStop}%
\bibitem [{\citenamefont {Palaniappan}\ \emph {et~al.}(2023)\citenamefont
  {Palaniappan}, \citenamefont {Ramanan},\ and\ \citenamefont
  {Urban}}]{Palaniappan2023}%
  \BibitemOpen
  \bibfield  {author} {\bibinfo {author} {\bibfnamefont {V.}~\bibnamefont
  {Palaniappan}}, \bibinfo {author} {\bibfnamefont {S.}~\bibnamefont
  {Ramanan}},\ and\ \bibinfo {author} {\bibfnamefont {M.}~\bibnamefont
  {Urban}},\ }\bibfield  {title} {\bibinfo {title} {{Equation of state of
  superfluid neutron matter with low-momentum interactions}},\ }\href
  {https://doi.org/10.1103/PhysRevC.107.025804} {\bibfield  {journal} {\bibinfo
   {journal} {Phys. Rev. C}\ }\textbf {\bibinfo {volume} {107}},\ \bibinfo
  {pages} {025804} (\bibinfo {year} {2023})}\BibitemShut {NoStop}%
\bibitem [{\citenamefont {Palaniappan}\ \emph {et~al.}(2025)\citenamefont
  {Palaniappan}, \citenamefont {Ramanan},\ and\ \citenamefont
  {Urban}}]{Palaniappan2025}%
  \BibitemOpen
  \bibfield  {author} {\bibinfo {author} {\bibfnamefont {V.}~\bibnamefont
  {Palaniappan}}, \bibinfo {author} {\bibfnamefont {S.}~\bibnamefont
  {Ramanan}},\ and\ \bibinfo {author} {\bibfnamefont {M.}~\bibnamefont
  {Urban}},\ }\bibfield  {title} {\bibinfo {title} {{Induced three-neutron
  interactions with low cutoffs for dilute neutron matter}},\ }\href
  {https://doi.org/10.1103/PhysRevC.111.035803} {\bibfield  {journal} {\bibinfo
   {journal} {Phys. Rev. C}\ }\textbf {\bibinfo {volume} {111}},\ \bibinfo
  {pages} {035803} (\bibinfo {year} {2025})}\BibitemShut {NoStop}%
\bibitem [{\citenamefont {Tabakin}(1969)}]{Tabakin1969}%
  \BibitemOpen
  \bibfield  {author} {\bibinfo {author} {\bibfnamefont {F.}~\bibnamefont
  {Tabakin}},\ }\bibfield  {title} {\bibinfo {title} {{Inverse Scattering
  Problem for Separable Potentials}},\ }\href
  {https://doi.org/10.1103/PhysRev.177.1443} {\bibfield  {journal} {\bibinfo
  {journal} {Phys. Rev.}\ }\textbf {\bibinfo {volume} {177}},\ \bibinfo {pages}
  {1443} (\bibinfo {year} {1969})}\BibitemShut {NoStop}%
\bibitem [{\citenamefont {Schrieffer}(1964)}]{Schrieffer}%
  \BibitemOpen
  \bibfield  {author} {\bibinfo {author} {\bibfnamefont {J.~R.}\ \bibnamefont
  {Schrieffer}},\ }\href@noop {} {\emph {\bibinfo {title} {{Theory of
  Superconductivity}}}}\ (\bibinfo  {publisher} {W. A. Benjamin},\ \bibinfo
  {address} {New York},\ \bibinfo {year} {1964})\BibitemShut {NoStop}%
\bibitem [{\citenamefont {Som\`a}\ \emph {et~al.}(2011)\citenamefont {Som\`a},
  \citenamefont {Duguet},\ and\ \citenamefont {Barbieri}}]{Soma2011}%
  \BibitemOpen
  \bibfield  {author} {\bibinfo {author} {\bibfnamefont {V.}~\bibnamefont
  {Som\`a}}, \bibinfo {author} {\bibfnamefont {T.}~\bibnamefont {Duguet}},\
  and\ \bibinfo {author} {\bibfnamefont {C.}~\bibnamefont {Barbieri}},\
  }\bibfield  {title} {\bibinfo {title} {{Ab initio self-consistent
  Gorkov-Green's function calculations of semimagic nuclei: Formalism at second
  order with a two-nucleon interaction}},\ }\href
  {https://doi.org/10.1103/PhysRevC.84.064317} {\bibfield  {journal} {\bibinfo
  {journal} {Phys. Rev. C}\ }\textbf {\bibinfo {volume} {84}},\ \bibinfo
  {pages} {064317} (\bibinfo {year} {2011})}\BibitemShut {NoStop}%
\bibitem [{\citenamefont {Perali}\ \emph {et~al.}(2002)\citenamefont {Perali},
  \citenamefont {Pieri}, \citenamefont {Strinati},\ and\ \citenamefont
  {Castellani}}]{Perali2002}%
  \BibitemOpen
  \bibfield  {author} {\bibinfo {author} {\bibfnamefont {A.}~\bibnamefont
  {Perali}}, \bibinfo {author} {\bibfnamefont {P.}~\bibnamefont {Pieri}},
  \bibinfo {author} {\bibfnamefont {G.~C.}\ \bibnamefont {Strinati}},\ and\
  \bibinfo {author} {\bibfnamefont {C.}~\bibnamefont {Castellani}},\ }\bibfield
   {title} {\bibinfo {title} {{Pseudogap and spectral function from
  superconducting fluctuations to the bosonic limit}},\ }\href
  {https://doi.org/10.1103/PhysRevB.66.024510} {\bibfield  {journal} {\bibinfo
  {journal} {Phys. Rev. B}\ }\textbf {\bibinfo {volume} {66}},\ \bibinfo
  {pages} {024510} (\bibinfo {year} {2002})}\BibitemShut {NoStop}%
\bibitem [{\citenamefont {Schirotzek}\ \emph {et~al.}(2008)\citenamefont
  {Schirotzek}, \citenamefont {Shin}, \citenamefont {Schunck},\ and\
  \citenamefont {Ketterle}}]{Schirotzek2008}%
  \BibitemOpen
  \bibfield  {author} {\bibinfo {author} {\bibfnamefont {A.}~\bibnamefont
  {Schirotzek}}, \bibinfo {author} {\bibfnamefont {Y.-i.}\ \bibnamefont
  {Shin}}, \bibinfo {author} {\bibfnamefont {C.~H.}\ \bibnamefont {Schunck}},\
  and\ \bibinfo {author} {\bibfnamefont {W.}~\bibnamefont {Ketterle}},\
  }\bibfield  {title} {\bibinfo {title} {{Determination of the Superfluid Gap
  in Atomic Fermi Gases by Quasiparticle Spectroscopy}},\ }\href
  {https://doi.org/10.1103/PhysRevLett.101.140403} {\bibfield  {journal}
  {\bibinfo  {journal} {Phys. Rev. Lett.}\ }\textbf {\bibinfo {volume} {101}},\
  \bibinfo {pages} {140403} (\bibinfo {year} {2008})}\BibitemShut {NoStop}%
\bibitem [{\citenamefont {Pieri}\ \emph {et~al.}(2004)\citenamefont {Pieri},
  \citenamefont {Pisani},\ and\ \citenamefont {Strinati}}]{Pieri2004}%
  \BibitemOpen
  \bibfield  {author} {\bibinfo {author} {\bibfnamefont {P.}~\bibnamefont
  {Pieri}}, \bibinfo {author} {\bibfnamefont {L.}~\bibnamefont {Pisani}},\ and\
  \bibinfo {author} {\bibfnamefont {G.~C.}\ \bibnamefont {Strinati}},\
  }\bibfield  {title} {\bibinfo {title} {{BCS-BEC crossover at finite
  temperature in the broken-symmetry phase}},\ }\href
  {https://doi.org/10.1103/PhysRevB.70.094508} {\bibfield  {journal} {\bibinfo
  {journal} {Phys. Rev. B}\ }\textbf {\bibinfo {volume} {70}},\ \bibinfo
  {pages} {094508} (\bibinfo {year} {2004})}\BibitemShut {NoStop}%
\bibitem [{\citenamefont {Haussmann}\ \emph {et~al.}(2007)\citenamefont
  {Haussmann}, \citenamefont {Rantner}, \citenamefont {Cerrito},\ and\
  \citenamefont {Zwerger}}]{Haussmann2007}%
  \BibitemOpen
  \bibfield  {author} {\bibinfo {author} {\bibfnamefont {R.}~\bibnamefont
  {Haussmann}}, \bibinfo {author} {\bibfnamefont {W.}~\bibnamefont {Rantner}},
  \bibinfo {author} {\bibfnamefont {S.}~\bibnamefont {Cerrito}},\ and\ \bibinfo
  {author} {\bibfnamefont {W.}~\bibnamefont {Zwerger}},\ }\bibfield  {title}
  {\bibinfo {title} {{Thermodynamics of the BCS-BEC crossover}},\ }\href
  {https://doi.org/10.1103/PhysRevA.75.023610} {\bibfield  {journal} {\bibinfo
  {journal} {Phys. Rev. A}\ }\textbf {\bibinfo {volume} {75}},\ \bibinfo
  {pages} {023610} (\bibinfo {year} {2007})}\BibitemShut {NoStop}%
\bibitem [{\citenamefont {Bogner}\ \emph {et~al.}(2010)\citenamefont {Bogner},
  \citenamefont {Furnstahl},\ and\ \citenamefont {Schwenk}}]{Bogner2010}%
  \BibitemOpen
  \bibfield  {author} {\bibinfo {author} {\bibfnamefont {S.~K.}\ \bibnamefont
  {Bogner}}, \bibinfo {author} {\bibfnamefont {R.~J.}\ \bibnamefont
  {Furnstahl}},\ and\ \bibinfo {author} {\bibfnamefont {A.}~\bibnamefont
  {Schwenk}},\ }\bibfield  {title} {\bibinfo {title} {{From low-momentum
  interactions to nuclear structure}},\ }\href
  {https://doi.org/10.1016/j.ppnp.2010.03.001} {\bibfield  {journal} {\bibinfo
  {journal} {Prog. Part. Nucl. Phys.}\ }\textbf {\bibinfo {volume} {65}},\
  \bibinfo {pages} {94} (\bibinfo {year} {2010})}\BibitemShut {NoStop}%
\bibitem [{\citenamefont {Ramanan}\ and\ \citenamefont
  {Urban}(2013)}]{Ramanan2013}%
  \BibitemOpen
  \bibfield  {author} {\bibinfo {author} {\bibfnamefont {S.}~\bibnamefont
  {Ramanan}}\ and\ \bibinfo {author} {\bibfnamefont {M.}~\bibnamefont
  {Urban}},\ }\bibfield  {title} {\bibinfo {title} {{BEC-BCS Crossover in
  Neutron Matter with Renormalization Group based Effective Interactions}},\
  }\href {https://doi.org/10.1103/PhysRevC.88.054315} {\bibfield  {journal}
  {\bibinfo  {journal} {Phys. Rev. C}\ }\textbf {\bibinfo {volume} {88}},\
  \bibinfo {pages} {054315} (\bibinfo {year} {2013})}\BibitemShut {NoStop}%
\bibitem [{\citenamefont {Ramanan}\ and\ \citenamefont
  {Urban}(2021)}]{Ramanan2021}%
  \BibitemOpen
  \bibfield  {author} {\bibinfo {author} {\bibfnamefont {S.}~\bibnamefont
  {Ramanan}}\ and\ \bibinfo {author} {\bibfnamefont {M.}~\bibnamefont
  {Urban}},\ }\bibfield  {title} {\bibinfo {title} {{Pairing in pure neutron
  matter}},\ }\href {https://doi.org/10.1140/epjs/s11734-021-00008-0}
  {\bibfield  {journal} {\bibinfo  {journal} {Eur. Phys. J. ST}\ }\textbf
  {\bibinfo {volume} {230}},\ \bibinfo {pages} {567} (\bibinfo {year}
  {2021})}\BibitemShut {NoStop}%
\bibitem [{\citenamefont {Bulgac}\ \emph {et~al.}(2008)\citenamefont {Bulgac},
  \citenamefont {Drut},\ and\ \citenamefont {Magierski}}]{Bulgac2008}%
  \BibitemOpen
  \bibfield  {author} {\bibinfo {author} {\bibfnamefont {A.}~\bibnamefont
  {Bulgac}}, \bibinfo {author} {\bibfnamefont {J.~E.}\ \bibnamefont {Drut}},\
  and\ \bibinfo {author} {\bibfnamefont {P.}~\bibnamefont {Magierski}},\
  }\bibfield  {title} {\bibinfo {title} {{Quantum Monte Carlo simulations of
  the BCS-BEC crossover at finite temperature}},\ }\href
  {https://doi.org/10.1103/PhysRevA.78.023625} {\bibfield  {journal} {\bibinfo
  {journal} {Phys. Rev. A}\ }\textbf {\bibinfo {volume} {78}},\ \bibinfo
  {pages} {023625} (\bibinfo {year} {2008})}\BibitemShut {NoStop}%
\bibitem [{\citenamefont {Galitskii}(1958)}]{Galitskii1958}%
  \BibitemOpen
  \bibfield  {author} {\bibinfo {author} {\bibfnamefont {V.~M.}\ \bibnamefont
  {Galitskii}},\ }\bibfield  {title} {\bibinfo {title} {{The energy spectrum of
  a non-ideal Fermi gas}},\ }\href
  {http://jetp.ras.ru/cgi-bin/dn/e_007_01_0104.pdf} {\bibfield  {journal}
  {\bibinfo  {journal} {Sov. Phys. JETP}\ }\textbf {\bibinfo {volume} {34}},\
  \bibinfo {pages} {104} (\bibinfo {year} {1958})}\BibitemShut {NoStop}%
\bibitem [{\citenamefont {Fetter}\ and\ \citenamefont
  {Walecka}(1971)}]{FetterWalecka}%
  \BibitemOpen
  \bibfield  {author} {\bibinfo {author} {\bibfnamefont {A.~L.}\ \bibnamefont
  {Fetter}}\ and\ \bibinfo {author} {\bibfnamefont {J.~D.}\ \bibnamefont
  {Walecka}},\ }\href@noop {} {\emph {\bibinfo {title} {{Quantum Theory of
  Many-Particle Systems}}}}\ (\bibinfo  {publisher} {McGraw-Hill},\ \bibinfo
  {address} {New York},\ \bibinfo {year} {1971})\BibitemShut {NoStop}%
\bibitem [{\citenamefont {Wellenhofer}\ \emph {et~al.}(2021)\citenamefont
  {Wellenhofer}, \citenamefont {Drischler},\ and\ \citenamefont
  {Schwenk}}]{Wellenhofer2021}%
  \BibitemOpen
  \bibfield  {author} {\bibinfo {author} {\bibfnamefont {C.}~\bibnamefont
  {Wellenhofer}}, \bibinfo {author} {\bibfnamefont {C.}~\bibnamefont
  {Drischler}},\ and\ \bibinfo {author} {\bibfnamefont {A.}~\bibnamefont
  {Schwenk}},\ }\bibfield  {title} {\bibinfo {title} {{Effective field theory
  for dilute Fermi systems at fourth order}},\ }\href
  {https://doi.org/10.1103/PhysRevC.104.014003} {\bibfield  {journal} {\bibinfo
   {journal} {Phys. Rev. C}\ }\textbf {\bibinfo {volume} {104}},\ \bibinfo
  {pages} {014003} (\bibinfo {year} {2021})}\BibitemShut {NoStop}%
\bibitem [{\citenamefont {Tinkham}(1975)}]{Tinkham}%
  \BibitemOpen
  \bibfield  {author} {\bibinfo {author} {\bibfnamefont {M.}~\bibnamefont
  {Tinkham}},\ }\href@noop {} {\emph {\bibinfo {title} {Introduction to
  superconductivity}}}\ (\bibinfo  {publisher} {McGraw-Hill},\ \bibinfo {year}
  {1975})\BibitemShut {NoStop}%
\bibitem [{\citenamefont {Urban}\ and\ \citenamefont
  {Ramanan}(2026)}]{Urban2026}%
  \BibitemOpen
  \bibfield  {author} {\bibinfo {author} {\bibfnamefont {M.}~\bibnamefont
  {Urban}}\ and\ \bibinfo {author} {\bibfnamefont {S.}~\bibnamefont
  {Ramanan}},\ }\bibfield  {title} {\bibinfo {title} {{Data for ``Hartree shift
  and pairing gap in ultracold Fermi gases in the framework of low-momentum
  interactions'', arXiv:2602.17420}},\ }\href
  {https://doi.org/10.5281/zenodo.18710220} {10.5281/zenodo.18710220} (\bibinfo
  {year} {2026})\BibitemShut {NoStop}%
\end{thebibliography}%

\end{document}